\documentclass[12pt]{iiscthes}

\usepackage[T1]{fontenc}
\usepackage{latexsym}
\usepackage{amssymb}
\usepackage{amsmath}
\usepackage{float}
\usepackage{mathptmx}
\usepackage{bm}
\usepackage{url}
\usepackage[titletoc]{appendix}

\usepackage[final]{microtype}
\usepackage{subfig}
\usepackage{epsf}
\usepackage{amssymb}
\usepackage{graphicx}
\usepackage{amsmath}
\usepackage[english]{babel}
\usepackage{mathrsfs}
\usepackage{fancyhdr}
\usepackage{verbatim}

\usepackage{amsthm}
\usepackage{latexsym}
\usepackage{amssymb}
\usepackage{amsfonts}
\usepackage{amsmath}
\usepackage{hyperref}
\usepackage{subfig}
\usepackage{epsfig}
\usepackage{amsbsy}
\usepackage{bm}
\usepackage{caption}
\theoremstyle{definition}

\theoremstyle{definition}

\theoremstyle{remark}

\DeclareMathAlphabet{\mathpzc}{OT1}{pzc}{m}{it}

\usepackage{epsf}
\usepackage{amssymb}
\usepackage{graphicx}
\usepackage{stmaryrd}
\usepackage{amsmath}
\usepackage{mathrsfs}

\begin{document}

\begin{frontmatter}
%
%

\title{Effect of particles on spinodal decomposition: \\
A phase field study
}

\author{Supriyo Ghosh\\
\textbf{}\\
\textnormal{under the guidance of}\\
\textbf{}\\
\large{Prof. T. A. Abinandanan}
}

\submitdate{JUNE 2012}
\dept{Department of Materials Engineering}
\me
\iisclogotrue 
\maketitle
%
%
%

\prefacesection{Acknowledgements}
I would like to thank my advisor, Prof. T. A. Abinandanan, for his guidance and encouragement throughout my work. I cherish his motivating words and advice on my future career. I could not ask for a better advisor. I thank Dr. Suryasarathi Bose for his invaluable comments, suggestions and advise on my project. The entire faculty, staff and students in the Materials Engineering Department made my learning enjoyable. I would also like to thank my family, friends, and lab mates - Naveen, Bhaskar, Chaitanya and Vinay for their support and understanding. Finally, special thanks to Rooparam, Nikhil, Santanu, Ratul, Rajdipda, Samratda, Arkoda for their useful suggestions.


 \begin{abstract}
 The present work is directed towards the understanding of the interplay of phase separation and wetting which dominates the morphological evolution in multicomponent systems. For this purpose, we have studied the phase separation pattern of a binary mixture (AB) in presence of stationary spherical particles (C) which prefers one of the components of the binary (say, A). Binary AB is composed of critical composition(50:50) and off-critical compositions(60:40, 40:60). Off-critical compositions are chosen to include two cases where either the major or minor component wets the particle. Particles are fixed in position and spherical in shape. Particle sizes of 8 units and 16 units are used in all simulations. Two types of particle loading are used, 5\% and 10\%. Particles are well-distributed in the matrix at a certain interparticle distance following periodic boundary conditions. 
 
We have employed a ternary form of Cahn-Hilliard equation to model such system. This model is a modification of Bhattacharya's model to incorporate immobile fillers. Free energy of such an inhomogeneous system depends on both composition and composition gradients. Composition provides homogeneous contribution to the system free energy whereas composition gradients contribute to the interfacial energies. Homogeneous form of free energy is given by regular solution expression which is very closely related to Flory-Huggins model for monodisperse polymer mixtures. To elucidate the effect of wetting on phase separation we have designed three sets of $\chi_{ij}$ and $\kappa_{ij}$ to include the effects of neutral preference, weak preference and strong preference of the particle for one of the binary components. We have simulated two different cases where the binary matrix (A:B) is quenched critically or off-critically in presence of stationary spherical particles.   

If the particles are preferentially wetted by one of the components then early stage microstructures show transient concentric alternate layers of preferred and non-preferred phases around the particles. When particles are neutral to binary components then such a ring pattern does not form. At late times neutral preference between particles and binary components yields a cocontinuous morphology whereas preferential wetting produces isolated domains of non-preferred phases dispersed in a continuous matrix of preferred phase. In other words lack of preference forms a nearly complete phase separate morphology for a binary of critical composition whereas an incomplete phase separation is seen if preference exists between particle and matrix components. In all the cases the binary interaction parameters are such that $\chi_{AB} > |\chi_{BC} - \chi_{AC}|$, which refers to a equilibrium wetting state where particles are in contact with both the components with a surplus of preferred component around it. Particles at the interface provide a resistance to interfacial motion and thus impede domain coarsening. In addition, higher particle loading and smaller particle size are also highly effective in reducing the kinetics of phase separation and domain growth.  

For off-critical compositions we have studied two different situations where either major or the minor component wets the network. When minor component wets the particle then a bicontinuous morphology results whereas when major component wets the network a droplet morphology is seen. In such cases early stage morphology suggests an enriched layer of preferred component around the particle though it is fundamentally different than the "target" pattern formed in case of critical mixture. When majority component wets the particle, a possibility of double phase separation is reported. In such alloys phase separation starts near the particle surface and propagates to the bulk at intermediate to late times forming spherical or nearly spherical droplets of the minor component. 
 \end{abstract}

\makecontents
\listoftables


\vspace{10MM}

\noindent

   
\end{frontmatter}

\chapter{Introduction}
Consider a binary AB, either a polymer blend or a fluid mixture, which is homogeneous at high temperature. If this unstable or metastable mixture is quenched below the co-existing curve, it will thrust into  A-rich and B-rich domains. This phenomena is called phase separation. In addition, it is also possible that same AB binary can phase separate in presence of a surface with a preferential attraction to one of the components. This is called preferential wetting and it results into a partially or completely wetted surface by the preferred component. Thus, the interplay of these two kinetic processes, phase separation and wetting, produce a great richness of microstructures. This class of microstructures are of great technological, experimental and theoretical importance. 
\\ \\
Polymer materials are hardly ever used in their pure form in applications~\cite{Karim}.They are often filled with solid additives which dramatically improve the mechanical, thermal and interfacial properties of the material relative to the pure polymer blend~\cite{Balazs}. For example, rubber particle increases toughness, carbon black/flake/tube improves conductivity and processibility, silica/glass beads or fibers enhance modulus and strength, clay sheets modify the heat resistance of the matrix~\cite{Balazs} etc. Moreover, wetting induced phase separation can lead us to gain a novel composite structure of alternating domains of polymeric and metallic materials~\cite{Balazs}. In addition, fixed particles in a matrix induces a pinning effect and thus dictates the final domain size and distribution (bicontinuous or isolated)~\cite{Tanaka}. All these ideas could be applied to physical design of multilayer composites including polymer blends and polymer-dispersed liquid crystal displays~\cite{Tanaka}, thin films~\cite{RNauman}, super conductors~\cite{Link2}, shape memory alloys~\cite{Link2} and even nanotubes~\cite{Fan}. 
\\ \\
While phase separation in binary systems have been studied extensively, little is known about phase separation in ternary systems~\cite{Chen,Eyre,RNauman}. When, along with phase separation another kinetic process, preferential wetting, comes into picture then the problem becomes complex. Moreover, quantitative simulation of such situations need to consider melt condition thermodynamics (for polymer mixture) or hydrodynamics (for fluid mixture). This complicates the problem even further. However, few studies have shed some light on wetting induced phase separation where wetting surface is provided by stationary wall~\cite{Puri,Brown}, substrate~\cite{Puri}, spherical particle~\cite{Lee}, network~\cite{Chakrabarti}, pattern substrate~\cite{Puri} etc. Majority of research in this direction employed a Cahn-Hilliard-Cook approach to simulate phase separation of a critical mixture, via Ginzburg-Landau functional in conjunction with a velocity field to the surface. Phase separation of a off-critical binary mixture in presence of a surface also has not received much attention except only by few~\cite{Puri,Puri_Frisch,PBinder,Brown}. Hence, coupled wetting and phase separation of a critical/off-critical binary mixture in presence of a immobile spherical (symmetric to both components) particle can still be considered a new problem.
\\ \\
Our objective is to investigate the phase separation behavior of a critical/off-critical binary mixture in presence of fixed particles of variable size and density. For this purpose, we adopt a ternary spinodal model developed by Bhattacharya~\cite{Bhattacharya} and modified it to simulate such situation. This model is based on Cahn-Hilliard formalism~\cite{Cahn} where bulk free energy (regular solution model) is supplemented with a gradient squared term (gradient energy). We vary the pairwise interaction energy and gradient energy parameters to incorporate preferential attraction (wetting) between the particles and one of the components of the binary mixture. Simulations are carried out by semi-implicit numerical integration of time-dependent Cahn-Hilliard equation on a $512^2$ square lattice in Fourier space, subject to periodic boundary conditions in both x and y directions.
\\ \\
The present report is organized as follows. In chapter 2, we review the available experimental and theoretical literatures related to our study.
Chapter 3 contains detailed description of the numerical model and other supplementary numerical calculations regarding ternary phase separation. Chapter 4 deals with the background details of our simulation including parameter estimations and the modifications adopted to incorporate immobile filler particles in a modified Cahn-Hilliard model. In chapter 5, we present the results and discuss the possible mechanisms of the microstructural features. Finally, we conclude with our findings in chapter 6. 
\chapter{Literature Review}
In this chapter we review the experimental and numerical studies relevant to the phase separation of a binary mixture 
in presence of third component. Third component may be present in the matrix as filler, particle, tube, network, wall etc. and behaves as mobile or immobile. It can be scaled as microscopic or nanoscopic. It may even induce preferential wetting to one of the components of the binary mixture. Moreover, the binary mixture can be comprising of a polymer blend or a fluid mixture. All this variables leads to a multiplicity of interesting microstructures. Thus, a symbiotic interaction between experimental, theoretical and numerical studies are very important for development in this class.
\section{Experimental Studies}
While phase separation in binary mixture has been studied extensively for the past two decades, available literature is few on the same in presence of solid particles. Amongst them most of the paper reports about the phase-separation morphology of binary mixture having critical composition. So, off-critical phase separation behavior has received very less attention so far.
\\ \\
The first related experimental study was conducted by Jones \emph{et. al.}~\cite{Puri}. They studied the spinodal decomposition behavior of a critical binary polymer mixture of poly(ethylene-propylene) (PEP) and perdeuterated-PEP (dPEP) in a presence of surface which preferentially attracts dPEP. They reported the origin of concentration waves from the surface. Another connected study is found in Bruder \emph{et. al.}~\cite{Puri} and Straub \emph{et. al.}~\cite{Puri}. They have used critical mixture of deutarated polystyrene (dPS) and brominated polystyrene (PBr$_x$S). In this situation preferential wetting resulted in a metastable CW morphology, which finally decomposed into a PW morphology. 
\\ \\
Tamai, Tran-cong and others~\cite{Lee} also experimented using PS/PVME with a crosslink-able side group styrene-chloromethyl styrene random copolymer (PSCMS). The morphology exhibits a ring pattern of alternate layers of preferred and non-preferred phases. A similar situation was studied by Ermi, Karim and others~\cite{Lee}. Karim \emph{et. al.}~\cite{Karim} examined with PS/PVME polymer blend with immobile macroscopic silica beads with preferential interaction with PS. During intermediate stage, AFM images show a enriched PS composition layer about the filler. This pattern is also ring alike, which is termed as "target pattern". Breaking of the translational phase separation symmetry by the filler particles is the reason reported for such circular composition waves. Moreover, these "target patterns" are transient in nature and disintegrate at late times. The authors also shed some light on slowing down of the phase separation process due to the interfacial segregation of particles.     
\\ \\
To the best of our knowledge, Tanaka's observations~\cite{RTanaka} are most closely related with our results. Binary fluid mixtures was comprised of oligomers of styrene (OS) and $\epsilon$-caprolactone (OCL) and spherical macroscopic glass particles are used as spacer as well as immobile filler. The glass particles were sandwitched between two glass plates so that particles become essentially immobile due to the large friction against the glass plates. As a result during morphology evolution the more preferable OCL-rich phase forms domains around the glass particles and the glass particles, which are close enough, are essentially bridged by it. He also mentioned that coarsening of droplets  completely stops due to pinning of the same by the fixed glass particles. He also conducted same experiment with mobile particles. This also concluded with similar results including bridge pattern formation and spontaneous pinning by mobile particles~\cite{Tanaka}.
\\ \\
Another related experiment was performed by Benderly \emph{et. al.}~\cite{Benderly}.The system comprised binary mixture (off-critical) of polypropylene (PP) and polyamide-6 (PA -6) along with macroscopic glass beads/fibers as filler with preferential attraction to PP. SEM images show an encapsulated morphology when PP is the minor phase and there was no encapsulation reported when PP is major phase. Benderly \emph{et. al.}~\cite{Benderly} have done another set of experiment with PP/PC/glass blend. They predicted an encapsulated morphology about PC. But, they observed a different morphology (separately dispersed) and they blamed the kinetic factors like viscosity etc., for the kinetic hindrance to encapsulation. 
\section{Theoretical Studies}
Majority of the related theoretical modelling literature deals with the phase separation behavior of a critical binary mixture with presence of solid additives with preferential attraction to one of the components of that mixture. However, available literature (known to us) regarding off-critical phase separation is very few ~\cite{PBinder,Brown,Chakrabarti,Ma}. Filler surfaces are introduced in the models as stationary wall~\cite{Puri,Brown}, substrate~\cite{Puri}, sphere~\cite{Lee}, network~\cite{Chakrabarti} etc. Various groups have simulated with patterned substrates like checkerboard or strip pattern so that different regions have different interaction with the third component~\cite{Puri}. 
\\ \\
The experiment of Karim \emph{et. al.}~\cite{Karim} was motivated by theoretical study of  Lee \emph{et. al.}~\cite{Lee}. Scope of our simulation seems to be most closely related with the same. They adapted a Cahn-Hilliard-Cook (CHC) model and the free energy functional considered was Ginzburg-Landau (GL) form subject to an initial condition of random thermal noise (white noise). The put a isolated spherical immobile particle in the center of $ 128^2 $ matrix, and allowed the A/B critical binary to phase separate with the constraints of variable polymer-filler interaction (controlled by parameter h). In case of strong interaction the early stage microstructures suggest a concentric ring pattern of alternate layers of wetting phase and non-wetting phase, whereas neutral interaction between polymer-filler does not produce any "target pattern". Moreover, the authors reported that such filler induced composition wave is transient and breaks up when the background spinodal pattern coarsens to a scale larger than filler particle. They also accounted an off-critical composition and claimed that if minor phase is the preferred phase then an encapsulated layer forms around filler particle though it can not be considered as "target pattern".   
\\ \\
A recent simulation  study by Hore~\emph{et. al.}~\cite{Hore} claimed that nanoparticles segregate at the interfaces if the mutual pairwise interaction parameters are such that $\chi_{AB} > |\chi_{BC} - \chi_{AC}|$. Their simulation was motivated by an experiment of Chung~\emph{et. al.}[Nano Lett. 5, 1878 (2005)]. The model was based on Dissipative particle dynamics (DPD) and a rigid body dynamics was employed to impart the velocity field to the spherical nanoparticles. Moreover, they reported that domain growth of a A/B binary decreases if the volume fraction of the particles increases and radius of the same decreases.
\\ \\
A series of simulations by Balazs~\emph{et. al.}~\cite{ACBalazs} shed light on this situation~\cite{DCBalazs,FCBalazs}. They focus on both issues of fixed particle~\cite{DCBalazs,FCBalazs} and mobile particle~\cite{ACBalazs} which was subject to selective interaction to one of the components in a  critical binary mixture. The authors used Cahn-Hilliard type approach, where free energy functional was considered as GL form along with a coupling contribution to incorporate the interaction between particles and the order parameter field. Navier-Stokes equation or Langevin dynamics were employed to incorporate hydrodynamic effects or particle mobility. All these simulations reported late stage morphology when particles sit at the interface and act as an obstacle to interface motion resulting a distribution of non-wettable phase as isolated islands in a continuous sea of wettable phase.     
\\ \\
Chakrabarti~\cite{Chakrabarti}, Brown~\cite{Brown,BAChakrabarti} have also used similar models to probe the surface directed spinodal decomposition~\cite{BAChakrabarti} and surface directed nucleation~\cite{Chakrabarti,Brown}. They modeled off-critical bulk phase separation by a CHC approach via GL functional in conjuction with a surface potential term or long range interaction term. They used fumed silica network~\cite{Chakrabarti} or block coplymer of strip pattern~\cite{Brown} to study the phase separation. Following them, if major component wets the surface then minor droplets nucleates near the surface before they nucleates in bulk. They reported a late time arrested growth of the wetting layer in case of minor component wets the network. Another similar numerical study is due to puri and Binder~\cite{PBinder}. They simulated bulk off-critical A/B phase separation with preferential attraction to a stationary surface. Following them, if minor component wets the surface, droplets of wetting components form whereas in case of major component wets the network non-wetting component form the droplets. They also reported a enriched layer of preferred component at the surface, in case of minor component wets the surface.  
\\ \\
Finally, we will conclude this chapter with a beautiful review paper of Nauman and He~\cite{RNauman}. This review discusses their contribution to this stimulating field in pedagogical framework. Here, the authors talk about the non-ideal diffusion in small and large binary/ternary systems. They considered enthalpic and entropic contribution to the gradient energy parameter, and supplied a pre-gradient mole fraction in flux equation. All these result a modified Cahn-Hilliard equation which leads to a better agreement with experiments. According to them, white noise (initial condition of random noise) can be considered as reasonable estimate to thermal noise and magnitude of such noise does not have any significant effect in middle to late stage phase separation. We will reproduce some of the related microstructures from this review to establish qualitative agreement with our results.
\begin{figure}[H]
\begin{center}
\includegraphics[scale=0.4]{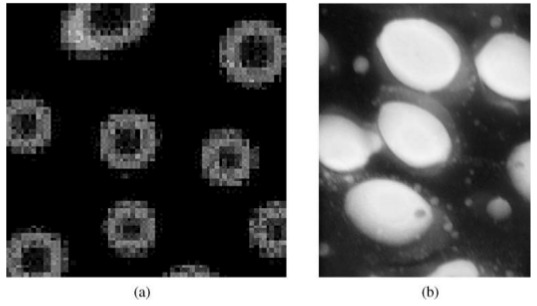}
\caption[Simulated (a) and experimental (b) core shell morphoology]{Simulated (a) and experimental (b) core shell morphoology (Reproduced from~\cite{RNauman})}\label{he1}
\end{center}
\end{figure}
\begin{figure}[H]
\begin{center}
\includegraphics[scale=0.4]{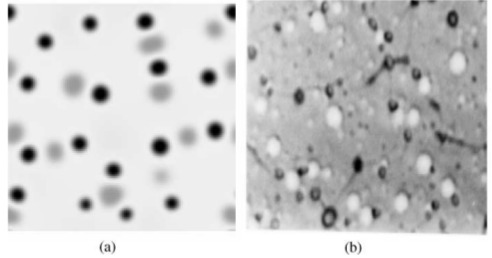}
\includegraphics[scale=0.4,height=3.7cm,width=3.7cm]{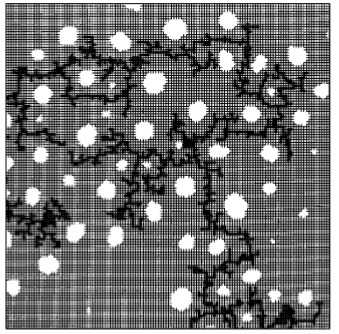}
\caption[Simulated (a) and experimental (b) droplet morphology]{Simulated (a) and experimental (b) droplet morphology (Reproduced from ~\cite{RNauman,Chakrabarti})(surface directed nucleation)}\label{he2}
\end{center}
\end{figure}
\begin{figure}[H]
\begin{center}
\includegraphics[scale=0.4]{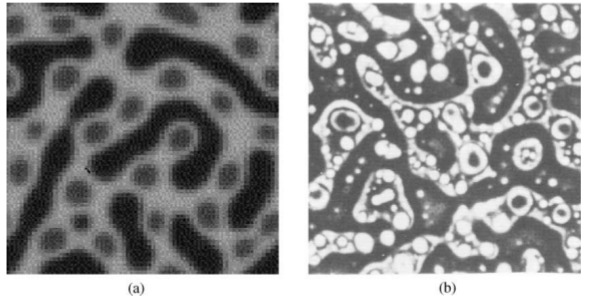}
\caption[Simulated (a) and experimental (b) core/shell morphology with continuous shell]{Simulated (a) and experimental (b) core/shell morphology with continuous shell (Reproduced from~\cite{RNauman})}\label{he2}
\end{center}
\end{figure}
\begin{figure}[H]
\begin{center}
\includegraphics[scale=0.4,width=4.0cm,height=4.0cm]{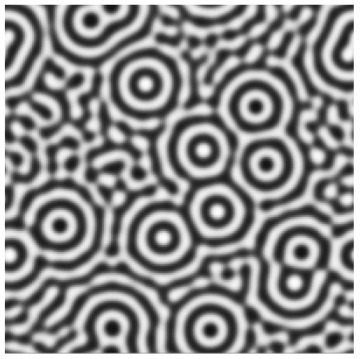}
\includegraphics[scale=0.4,width=4.0cm,height=4.0cm]{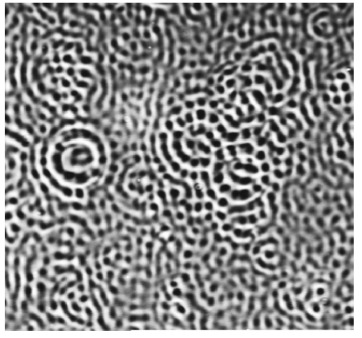}\\
\caption[Simulated (a) and experimental (b) ring pattern]{Simulated (a) and experimental (b) ring pattern (reproduced from~\cite{Lee})(surface directed spinodal decomposition)}\label{leese}
\end{center}
\end{figure}
\chapter{Formulation}
In this chapter we deal with the formulation for the dynamics of the spinodal decomposition in ternary alloys. At first we derive a free energy functional by the local free energy expression from regular solution model and the inhomogeneous free energy expression from Cahn-Hilliard model. Then sequentially with the help of continuity equation, Fourier transform and finite difference approach  we obtain the kinetic equations which governs the temporal evolution of composition field. We adopted the similar numerical model practiced by Bhattacharya~\cite{Bhattacharya}.
\section{Regular Solution Model}
We consider a ternary alloy system consisting of three different species A,B and C. Let  $c_i\left(\mathbf{r},t\right)$  for i = A, B, C represent the mole fraction of the `i'th component as a function of position \textbf{r} and time $t$. Since $c_i$ is the mole fraction we have the following conditions:
\begin{equation}\label{51}
\sum_{i=A,B,C}{c_i}\left(\mathbf{r},t\right) = 1
\end{equation}
from regular solution model the bulk chemical free energy $ f\left(c_A,c_B,c_C\right) $ is given by
\begin{equation}\label{52}
 f\left(c_A,c_B,c_C\right)=\frac{1}{2}\sum_{i\neq j}\chi_{ij}c_ic_j+\sum_ic_i\ln c_i
 \end{equation}
where i,j = A,B,C and $\chi_{ij} = \chi_{ji}$ is the effective interaction energy between components i and j. Here we considered only pair--wise interactions or nearest neighbour approximation between the atoms.
\begin{equation}\label{ie}
\chi_{AB} = \dfrac{Z\left[2E_{AB}-E_{AA}-E_{BB}\right]}{2K_B T}
\end{equation}
where $E_{AB},E_{AA},E_{BB}$ are the bond energies between A/B,A/A and B/B bonds respectively, Z is the number of bonds per atom, $K_B$ is the Boltzmann constant and T is the absolute temperature. If $\chi_{AB}\,>0$ then binary system A-B exhibits a miscibility gap and if $\chi_{AB}\,>0$ then the binary solution exhibits ordering.

\section{Flory--Huggins Model}
The Flory--Huggins model for polymer solutions is a close relative of the regular solution model. According to this model in case of an incompressible A-B-C polymer blends ($\phi_A$+$\phi_B$+$\phi_C$ = 1) :
\begin{equation}
\frac{f}{KT} = \frac{\phi_A \ln \phi_A}{N_A} +\frac{\phi_B \ln \phi_B}{N_B}+\frac{\phi_C \ln \phi_C}{N_C} + \chi_{AB}\phi_A \phi_B +\chi_{BC}\phi_B \phi_C +\chi_{AC}\phi_A \phi_C
\end{equation}
$f$ is the free energy of mixing of the ternary solution. $\phi_A$, $\phi_B$, $\phi_C$ are volume fractions and $N_A$, $N_B$, $N_C$ are degree of polymerization of components A, B, C respectively. Degree of polymerization depends on polymer chain length and the number of chain segments (\emph{mer} content). If it is assumed that $N_A = N_B = N_C = 1$, then this model resembles to regular solution model. Similar to regular solution model $\chi_{AB}$, $\chi_{BC}$, $\chi_{AC}$ are the effective interaction energy between A--B, B--C, A--C binaries. 
\section{The Cahn-Hilliard Model}
The Cahn-Hilliard model adds a correction to the homogeneous free energy function to account for spatial inhomogeneity.This correction comes from a Taylor expansion of $ f\left(c_A,c_B,c_C\right) $ in powers of $\nabla c$ combined with symmetry considerations. While composition in a homogeneous system is scalar, composition becomes a field for an inhomogeneous system.Thus the total free energy becomes a functional of the compositional field[\cite{Cahn} given by:
\begin{equation}\label{che} F=N_v\int_{V}\left[f\left(c_A,c_B,c_C\right)+\sum_{i=A,B,C}\kappa_i\left(\nabla c_i\right)^2\right]dV \end{equation}
where f is the homogeneous free energy,and $N_V$ is the number of sites per unit volume.$\kappa$ is called the gradient energy co-efficient. $\kappa_i\left(\nabla c_i\right)^2$ is the gradient energy which, is the first order correction for inhomogeneity, introduces a penalty for sharp gradients and make the interface a diffuse one. Volume fraction also can be treated similarly, instead of composition, as a conserved phase field variables to simulate a polymer solution.
\section{Chemical Potential}
According to Onsager relations, the flux of the $i$ element, $ \mathbf{J}_i $, is proportional to the gradient of the chemical potential
\begin{equation}\label{53}
\mathbf{J_i}\left(\mathbf{x},t\right)= -M_i\nabla \mu_i\left(\mathbf{x},t\right) 
\end{equation}
 where $ M_i $ is the onsager coefficient (mobility of i th component)  and is always positive. Since mobility is isotropic for cubic materials it may be replaced by a scalar instead of second rank property tensor as in the previous equation.\\
 \\
In constructing the kinetic equation of a substitutional alloy undergoing diffusion we adopt the results of Kramer \emph{et al.}~\cite{kramer}, who proposed that there must be a net vacancy flux operating during the diffusion process with the constraint of local thermal equilibrium of vacancies everywhere. Thus the net flux of component $i$, $\mathbf{\bar J}_i $, across a fixed lattice plane (not with respect to inert mobile markers) is the sum of the diffusion flux of A plus the A transported by the vacancy flux.
\begin{eqnarray}\label{v}
\mathbf{\bar J_i} &=& \mathbf{J_i} + c_i \mathbf{J_V}\nonumber\\
\mathbf{J_V} &=& -(\mathbf{J_A} + \mathbf{J_B} + \mathbf{J_C})
\end{eqnarray}

where $J_V$  is the vacancy flux.
Combining equations~\ref{v} we get : 
\begin{equation}\label{54}  
\mathbf{\bar J}_i = \mathbf{J}_i-c_i\sum_{i=A,B,C}\left(\mathbf{J}_i\right)
\end{equation}
from eqn.~\ref{54} and eqn.~\ref{51}, it is clear that
\begin{equation}\label{55}
\sum_{i=A,B,C}\left(\mathbf{\bar J}_i\right)= 0
\end{equation}  
so the net flux of each component A,B,C becomes:
\begin{eqnarray}\label{56}
\mathbf{\bar J}_A = -\left(1-c_A\right)M_A\nabla\mu_A + c_A M_B\nabla\mu_B + c_AM_C\nabla\mu_C\nonumber\\
\mathbf{\bar J}_B = -\left(1-c_B\right)M_B\nabla\mu_B + c_B M_A\nabla\mu_A + c_BM_C\nabla\mu_C\nonumber\\
\mathbf{\bar J}_C = -\left(1-c_C\right)M_C\nabla\mu_C + c_C M_A\nabla\mu_A + c_CM_B\nabla\mu_B
\end{eqnarray}

We need to solve for the only two compositional variables, say $ c_A$ and $c_B $ as other composition can be directly computed by substraction of $ c_A$ and $c_B $ from unity. Now we need to derive an expression for $ \mathbf{\bar J}_i $
by Gibbs-Duhem equation:
\begin{equation}\label{57}
c_A\nabla\mu_A +c_B\nabla\mu_B+c_C\nabla\mu_C = 0
\end{equation}
from equations ~\ref{51} and ~\ref{57}  and rearranging we get set of equations :
\begin{eqnarray}\label{58}
\nabla\mu_A &=& (1-c_A)\nabla\mu_A^{eff}-c_B\nabla\mu_B^{eff}\nonumber\\
\nabla\mu_B &=& (1-c_B)\nabla\mu_B^{eff}-c_A\nabla\mu_A^{eff}\nonumber\\
\nabla\mu_C &=& -c_A\nabla\mu_A^{eff}-c_B\nabla\mu_B^{eff}
\end{eqnarray}
\[ \mbox where{ \nabla\mu_A^{eff} = \nabla\mu_A-\nabla\mu_c\; \& \;\nabla\mu_B^{eff} =\nabla\mu_B-\nabla\mu_C }\]\\
using the expressions for $ \nabla\mu_A$,$ \nabla\mu_B$,$ \nabla\mu_C $ in eqn.~\ref{56} and arraging we can write:
\begin{eqnarray}\label{59}
\mathbf{\bar J}_A = -\left[\left(1-c_A\right)^2M_A+c_A^2\left(M_B+M_C\right)\right]\nabla\mu_A^{eff}+\left[c_BM_A\left(1-c_A\right)+c_AM_B\left(1-c_B\right)-c_Ac_BM_C\right]\nabla\mu_B^{eff}\nonumber\\
\end{eqnarray}
\begin{eqnarray}\label{60}
\mathbf{\bar J}_B = -\left[\left(1-c_B\right)^2M_B+c_B^2\left(M_A+M_C\right)\right]\nabla\mu_B^{eff}+\left[c_AM_B\left(1-c_B\right)+c_BM_A\left(1-c_A\right)-c_Ac_BM_C\right]\nabla\mu_B^{eff}\nonumber\\
\end{eqnarray}
Now let us define the effective mobilities:
\begin{eqnarray}\label{mobility}
M_{AA}&=&\left(1-c_A\right)^2M_A+c_A^2\left(M_B+M_C\right)\nonumber\\
M_{BB}&=&\left(1-c_B\right)^2M_B+c_B^2\left(M_A+M_C\right)\nonumber\\
M_{AB}=M_{BA}&=&\left(1-c_A\right)c_BM_A+c_AM_B\left(1-c_B\right)-c_Ac_BM_C
\end{eqnarray}
Using the relations in eqn.~\ref{mobility}, we can rewrite the flux equations ~\ref{59} and ~\ref{60} in a more compact form :
\begin{eqnarray}\label{61}
\mathbf{\bar J}_A = -M_{AA}\nabla\mu_A^{eff}+M_{AB}\nabla\mu_B^{eff}\nonumber\\
\mathbf{\bar J}_B = -M_{BB}\nabla\mu_B^{eff}+M_{AB}\nabla\mu_A^{eff}
\end{eqnarray}

Chemical potential in homogeneous system is proportional to the partial derivative of bulk free energy:
\begin{equation}\label{62}
\mu_A^{eff}=\frac{\partial f\left(c_A,c_B\right)}{\partial c_A}
\end{equation}
Chemical potential in a inhomogeneous system is away from global equilibrium. So if we assume local equilibrium, we can define this potential field by employing the calculus of variations:
\begin{equation}\label{63}
\mu_i^{eff}=\frac{\delta F}{\delta c_i}\hspace{1in} where,\; i = A,B
\end{equation}
The variational derivative may be found by applying the Euler-Lagrange equation:
\begin{equation}\label{64}
\frac{\delta F}{\delta c_i} = \frac{\partial F}{\partial c_i}-\nabla \cdot \frac{\partial F}{\partial \nabla c_i}
\end{equation}
So, we obtain the following equations:
\begin{eqnarray}\label{65}
\mu_A^{eff}&=&\frac{\partial f}{\partial c_A}-2\left(\kappa_A+\kappa_C\right)\nabla^2c_A - 2\kappa_C\nabla^2c_B\nonumber\\
\mu_B^{eff}&=&\frac{\partial f}{\partial c_B}-2\left(\kappa_B+\kappa_C\right)\nabla^2c_B-2\kappa_C\nabla^2 C_A
\end{eqnarray}
where,
\begin{eqnarray}\label{66}
\frac{\partial f}{\partial c_A}&=&\ln c_A-\ln c_C+\left(\chi_{AB}-\chi_{BC}\right)c_B+\chi_{AC}\left(c_C-c_A\right)\nonumber\\
\frac{\partial f}{\partial c_B}&=&\ln c_B-\ln c_C+\left(\chi_{AB}-\chi_{AC}\right)c_A+\chi_{BC}\left(c_C-c_B\right)
\end{eqnarray}
\section{Evolution Equations}
Since c is a conserved quantity, it obeys a conservative (continuity) law, which can be used to obtain the expressions for the temporal evolution of composition field:
\begin{equation}\label{67}
\frac{\partial c_i}{\partial t}= - \nabla\cdot\mathbf{\bar J}_i
\end{equation}
using equations ~\ref{61} and ~\ref{65} in equation ~\ref{67} we get the following two independent kinetic equations:
\begin{eqnarray}\label{e1}
\frac{\partial c_A}{\partial t}= M_{AA}\left[\nabla^2 \left(\frac{\partial f}{\partial c_A}\right)-2\left(\kappa_A+\kappa_C\right)\nabla^4 c_A - 2\kappa_C\nabla^4 c_B\right]\nonumber\\
-M_{AB}\left[\nabla^2 \left(\frac{\partial f}{\partial c_B}\right)-2\left(\kappa_B+\kappa_C\right)\nabla^4 c_B-2\kappa_C\nabla^4 c_A\right]
\end{eqnarray}
\begin{eqnarray}\label{e2}
\frac{\partial c_B}{\partial t}= M_{BB}\left[\nabla^2 \left(\frac{\partial f}{\partial c_B}\right)-2\left(\kappa_B+\kappa_C\right)\nabla^4 c_B - 2\kappa_C\nabla^4 c_A\right]\nonumber\\
-M_{AB}\left[\nabla^2 \left(\frac{\partial f}{\partial c_A}\right)-2\left(\kappa_A+\kappa_C\right)\nabla^4 c_A-2\kappa_C\nabla^4 c_B\right]
\end{eqnarray}
where, for the sake of convenience we denoted $ \kappa_{AA}=\kappa_A+\kappa_C ,\kappa_{BB}=\kappa_B+\kappa_C \& \kappa_{AB}=\kappa_{BA}=\kappa_C $ and mobility is considered independent of composition field.
\section{Numerical Implementation}
We extended the 2-D semi-implicit Fourier spectral method to the ternary systems and obtain a sequence of time dependent ordinary differential equation in Fourier space. If we consider a function $ g_A = \frac{\partial f}{\partial c_A}$ and $ g_B = \frac{\partial f}{\partial c_B}$ then the expression becomes:\\
\begin{eqnarray}\label{68}
\frac{\partial\tilde{c}_A\left(\mathbf{k},t\right)}{\partial t} = M_{AA}\left[-k^2\left(\frac{\partial f}{\partial c_A}\right)_\mathbf{k}-2\kappa_{AA} k^4\tilde{c}_A-2\kappa_{AB}k^4\tilde{c}_B\right]\nonumber\\
-M_{AB}\left[-k^2\left(\frac{\partial f}{\partial c_B}\right)_\mathbf{k}-2\kappa_{AB}k^4\tilde{c}_A-2\kappa_{BB}k^4\tilde{c}_B\right]
\end{eqnarray}
\begin{eqnarray}\label{69}
\frac{\partial\tilde{c}_B\left(\mathbf{k},t\right)}{\partial t} = M_{BB}\left[-k^2\left(\frac{\partial f}{\partial c_B}\right)_\mathbf{k}-2\kappa_{BB} k^4\tilde{c}_B-2\kappa_{AB}k^4\tilde{c}_A\right]\nonumber\\
-M_{AB}\left[-k^2\left(\frac{\partial f}{\partial c_A}\right)_\mathbf{k}-2\kappa_{AB}k^4\tilde{c}_B-2\kappa_{AA}k^4\tilde{c}_A\right]
\end{eqnarray}
\\
 where $ \mathbf{k}=\left(k_x,k_y\right) \mbox {is the reciprocal lattice vector:}\, k=|\mathbf{k}|; \, \tilde{c}_A\left(\mathbf{k},t\right) \mbox{and} \, \tilde{c}_B\left(\mathbf{k},t\right)$ are the Fourier transforms of the respective compositions in the real space.
 Using finite difference method for $\frac{\partial c_A}{\partial t}$ and $\frac{\partial c_B}{\partial t}$ we get the following equations:
 \begin{equation}\label{70}
  \frac{\partial c_i}{\partial t}=\dfrac{\tilde{c}_i\left(\mathbf{k},t+\Delta t\right)-\tilde{c}_i\left(\mathbf{k},t\right)}{\Delta t} 
  \end{equation}
  where i = A,B. We treated the linear terms, $\tilde{c}_A $ and $\tilde{c}_B $ implicitly and the non-linear terms, $\tilde{g}_A $ and $\tilde{g}_B $ are treated explicitly i.e. $\tilde{g}_i\left(\mathbf{k},t+\Delta t\right)=\tilde{g}_i\left(\mathbf{k},t\right)$. We solve the following equations iteratively to obtain the microstructures.
 \begin{eqnarray}\label{71}
\tilde{c}_A\left(\mathbf{k},t+\Delta t\right)=\dfrac{\tilde{c}_A\left(\mathbf{k},t\right)-k^2\Delta t\left[M_{AA}\tilde{g}_A\left(\mathbf{k},t\right)-M_{AB}\tilde{g}_B\left(\mathbf{k},t\right)\right]-2k^4\Delta t\tilde{c}_B\left(\mathbf{k},t\right)\left[M_{AA}\kappa_{AB}-M_{AB}\kappa_{BB}\right]}{1+2\Delta t k^4\left(M_{AA}\kappa_{AA}-M_{AB}\kappa_{AB}\right)}\nonumber\\
\end{eqnarray}
 \begin{eqnarray}\label{72}
\tilde{c}_B\left(\mathbf{k},t+\Delta t\right)=\dfrac{\tilde{c}_B\left(\mathbf{k},t\right)-k^2\Delta t\left[M_{BB}\tilde{g}_B\left(\mathbf{k},t\right)-M_{AB}\tilde{g}_A\left(\mathbf{k},t\right)\right]-2k^4\Delta t\tilde{c}_A\left(\mathbf{k},t\right)\left[M_{BB}\kappa_{AB}-M_{AB}\kappa_{AA}\right]}{1+2\Delta t k^4\left(M_{BB}\kappa_{BB}-M_{AB}\kappa_{AB}\right)}\nonumber\\
\end{eqnarray}

\chapter{Simulation Details}
In this chapter we deal with the different variables or factors which influence the microstructure obtained by phase field simulations. First we discuss about loading of the particles, their size distribution, interparticle distance and distribution of particles within the matrix. Then, we define three systems ($S_O, S_W, S_S$) which includes the parameters like pair--wise interaction coefficient ($\chi$) and gradient energy coefficient ($\kappa$). Finally, how the composition of the phases are distributed in the matrix and what are the constant mobility  values assigned to the components are described. 

\section{Particle Characteristics}
\subsection{Particle Loading}
Simulations are performed with two types of particle loading : 5\% and 10\%.
\subsection{Particle Size and shape}
Shape of the Particles are circular in 2D and spherical in 3D. In 2D, we have considered circles of two different radius : 8 unit and 16 unit. According to the volume fraction ($\approx$ area fraction) of the particles, we calculated the numbers of circles of each radius (8 or 16) that can be arranged in 512$\times$512 matrix.

\begin{table}[h]
\begin{center}
\begin{tabular}{|c|c|r|}
\hline
Area fraction & Radius (R) & No. of particles\\
\hline
 5\% &  8 & 65\\ 
\cline{2-3}
&16 & 16\\ 
\hline
10\% &  8 & 130 \\ \cline{2-3}
&16 & 33\\
\hline
\end{tabular}\caption{Particle size and number of particles}\label{ps}
\end{center}
\end{table}

\subsection{Interparticle Distance ($\lambda$)}
To obtain  a well dispersed spatial arrangement of the particles we assigned an interparticle distance of greater than 5$\times$R  in cases of 5\% loading and a distance of greater than 4$\times$R  in cases of 10\% loading.
In other words, in case of 5\% loading domain size of the particle (diameter) is greater than interparticle distance and in case of 10\% loading domain size of particle approximately resembles to the interparticle distance. 

\subsection{Positioning}
\begin{figure}[h]
\begin{center}
\includegraphics[scale=0.4]{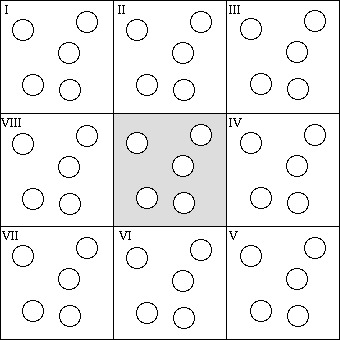}
\caption{Periodic boundary conditions}\label{pbc}
\end{center}
\end{figure}

particles are positioned at a interparticle distance in the matrix following periodic boundary conditions. This trick helps us to minimize the surface effects and also to simulate the properties of a system more closely. In periodic boundary conditions the cubical simulation box is replicated in all directions to form a infinite lattice. In the course of the simulation, when a molecule moves in the central box, its periodic image in every one of the other boxes moves with exactly the same orientation in exactly the same way. Thus, as a molecule leaves the central box, one of its images will enter through the opposite face. There are no walls at the boundary of the central box, and the system has no surface. The central box simply forms a convenient coordinate system for measuring locations of the N molecules~\cite{Link}.\\
Fig~\ref{pbc} is  a 2D version of such a periodic system. As a particle moves through a boundary, all its corresponding images move across their corresponding boundaries.
\section{Compositional Distribution}
It seems to be computationally more expensive when we assign the compositions to the particles of fixed radius (R), and  also to matrix and simulate the system to evolve. So we used the following trick to distributed the composition throughout the system. $c_p$ is the composition of the particles with is confined by the region of (R - dR), $c_m$ is composition of the matrix which occupies space at a distance greater than R + dR and compositions within the region in between them follows as a straight line relationship (refer~\ref{sl1},~\ref{sl2}). This approach gives us a c--rich phase (particle) of desired mean radius. All the simulations are performed keeping dR constant, 4 unit.
\[ 
c_i (r) = \left\{ 
\begin{array}{lr}
c_p & r < R - dR\\
straight\;line & R - dR < r < R + dR\\
c_m & r > R + dR
\end{array}
\right.
\]
\begin{equation}\label{sl1}
\frac{c_m - c_p}{2\,dR} = \frac{c - c_p}{r - (R - dR)}
\end{equation}
\begin{equation}\label{sl2}
c = c_p +\left[r - (R - dR)\right]\frac{c_m - c_p}{2 dR}
\end{equation}
\vspace{-20pt}
\begin{figure}[H]
\centering
\includegraphics[trim = 30mm 60mm 30mm 50mm,clip, scale=0.7]{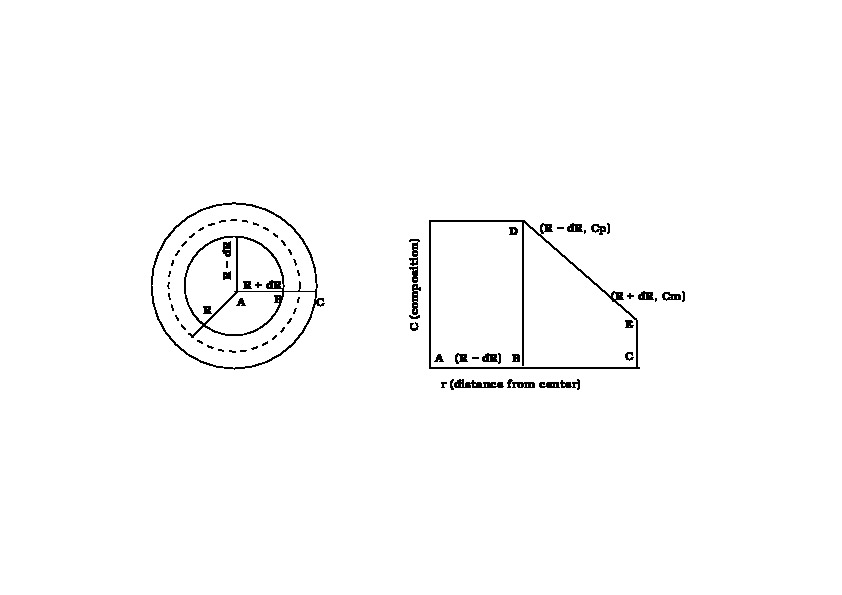}
\caption{Radial composition profile from the center of the particles (schematic)}
\label{cd}
\end{figure}
\section{Mobility Matrix}\label{mm}
In all our simulations particles are kept nearly immobile. Our objective is to show how these immobile particles affect the phase separation behaviour of the binary matrix phase. So,to achieve that we assign $M_c = 0$ in the equation~\ref{mobility} and it reduces to:
\begin{eqnarray}
M_{AA} &=& (1 - c_A)^2 M_A + c_A^2 M_B\nonumber\\
M_{BB} &=& (1 - c_B)^2 M_B + c_B^2 M_A\nonumber\\
M_{AB} &=& c_B (1 - c_A) M_A + c_A (1 - c_B) M_B
\end{eqnarray}
we judiciously selected the values of $c_A$ and $c_B$ with the following constraints : 
\begin{itemize}
\item determinant of mobility matrix must be positive definite 
\item it represents the matrix composition of an A--B binary mixture.
\item scaled mobilities of components A and B, $M_{AA}$ and $M_{BB}$ respectively, equals to one.  
\item every particle in the matrix survives.
\end{itemize}
Thus, the mobility matrix becomes the following and used throughout all simulations.
\begin{equation}
\textbf{M} =
\left[
\begin{array}{cc}
M_{AA}&M_{AB}\\
M_{AB}&M_{BB}
\end{array}
\right]
=
\left[
\begin{array}{cc}
1.0&0.98\\
0.98&1.0
\end{array}
\right]
\end{equation}
\section{Systems}
Wettability and phase separation behavior between components depends on mutual interaction  energy($\chi$) and interfacial energy ($\kappa$). we have used three distinct systems of different combination of $\chi$ and $\kappa$. Interaction energy is introduced in the system in terms of pairwise interaction parameters of regular solution model and gradient energy is integrated in terms of gradient energy coefficients of CH model. All this different variables will lead to three different ternary isothermal phase diagram from where we utilized the equilibrium compositions of three phases $\alpha$, $\beta$ and $\gamma$.
\subsection{Choice of Interaction Parameters}
Following equation~\ref{ie}, regular solution interaction parameter (per mole) can be written as : $\chi = \Omega/RT$ where,
\begin{equation}\Omega = \dfrac{Z\left[2E_{ij}-E_{ii}-E_{jj}\right]}{2} \end{equation}
$\chi$ is inversely proportional to critical temperature. So, instead of calculating a critical temperature for polymer solutions, we calculated $\chi_{crit}$, the critical value of interaction parameter at the onset of miscibility gap. 
\begin{equation}\label{cc}\chi > \chi_{crit}  = \Omega/RT_{crit}\end{equation}
It can be shown that the critical temperature (inflection point) in regular solution model is \begin{equation}\label{tc}
T_{crit} = \Omega / 2R
\end{equation}
Combining equations ~\ref{cc} and ~\ref{tc} we can conclude that at a value of $\chi$ greater than
2.0 i/j binary will phase separate.

Our objective was to preferentially wet the C rich particles by component A. So we intentionally attributed high interaction parameter values to B/C interface to make the system reluctant to form B/C interface in order to minimize its Gibbs free energy.
\begin{table}[h]
\begin{center}
\begin{tabular}{|c|c|c|c|}
\hline
$\chi$ & system-$S_O$ & system-$S_W$ & system-$S_S$\\
\hline
$\chi_{AB}$ & 2.5 & 2.5 & 2.5 \\
\hline\hline
$\chi_{BC}$ & 3.5 & 4.0 & 5.0 \\
\hline \hline
$\chi_{AC}$ & 3.5 & 3.5 & 3.5 \\
\hline
\end{tabular}
\end{center}
\end{table}
\subsection{Choice of Gradient Energy parameters} Our system is composed of A--rich and B--rich phases as a matrix  and C-rich phases (particles) are dispersed in between the A/B binary phase. Our focus was to preferentially wet the particles by one of the components (A). That's why we we assign higher interfacial energy to B/C interface in terms of larger gradient energy coefficient which eventually punishes the B--C interface more and creates a more diffuse interface between them.

From table~\ref{ok}, below, we can calculate three independent values of $\kappa_A$, $\kappa_B$, $\kappa_C$ for each type of system. Then we combine the corresponding values to form the computationally used parameters -- $\kappa_{AB}$, $\kappa_{BC}$, $\kappa_{AC}$.  
\begin{table}[h]
\begin{center}
\begin{tabular}{|c|c|c|c|}
\hline
$\kappa$ & system-$S_O$ & system-$S_W$ & system-$S_S$\\
\hline
$\kappa^{AB}$ = $\kappa_A$ + $\kappa_B$ & 8.0 & 8.0 & 8.0 \\
\hline
$\kappa^{BC}$ = $\kappa_B$ + $\kappa_C$ & 8.0 & 10.0 & 12.0 \\
\hline
$\kappa^{AC}$ = $\kappa_A$ + $\kappa_C$ & 8.0 & 8.0 & 8.0\\
\hline
\end{tabular}\caption{Theoretically considered values of $\kappa$}\label{ok}
\end{center}
\end{table}
\newpage  
\section{Synopsis of simulation parameters}\label{sp}
\begin{table}[h]
\begin{center}
 \begin{tabular}{|c|c|c|c|}
 \hline
 \textbf{Simulation parameters} & \textbf{system-$S_O$} & \textbf{system-$S_W$} & \textbf{system-$S_S$}\\
 \hline
 $ \Delta x $ & 1.0 &1.0&1.0\\
 \hline
 $ \Delta y $ & 1.0 &1.0&1.0\\
 \hline
 $ \Delta t $ & 0.005 &0.005&0.0005\\
 \hline
 System Size(x-dimension)  & 512 &512&512\\
 \hline
  System Size(y-dimension)  &512 & 512 &512\\
  \hline
  Composition fluctuation & $ \pm 0.005 $ & $\pm 0.005$ & $\pm 0.005$\\
  \hline\hline
  $ M_{AA} $ &1.0 & 1.0&1.0\\
  \hline
  $ M_{BB} $& 1.0 & 1.0&1.0\\
  \hline
  $ M_{AB} $ &0.98 & 0.98 & 0.98\\
  \hline\hline
  $ \kappa_ A $ & 3.0 & 2.0&1.0\\
  \hline
   $ \kappa _B $ & 4.0 & 5.0&6.0\\
  \hline
   $ \kappa _C $ & 4.0 & 5.0 & 6.0\\
  \hline \hline
   $ \kappa _{AA}=\kappa_A+\kappa_C $ & 8.0 & 8.0 & 8.0\\
  \hline
   $ \kappa _{BB}=\kappa_B+\kappa_C $ & 8.0 & 10.0 & 12.0\\
  \hline
   $ \kappa_ {AB}=\kappa_{BA}=\kappa_C $ & 4.0 & 5.0 & 6.0\\
\hline\hline
   $ \chi_{AB}$ & 2.5 & 2.5 & 2.5 \\
  \hline
   $ \chi _{BC}$ & 3.5 & 4.0 & 5.0 \\
  \hline
   $ \chi_ {AC}$ & 3.5 & 3.5 & 3.5\\   
  \hline\hline
  Matrix composition of A ($c_A^m$ ) &  0.45 & 0.45 &0.45\\
  \hline
  $c_B^m$ & 0.45 & 0.45 &0.45 \\
  \hline
  $c_C^m$ &  0.1 & 0.1 & 0.1 \\
  \hline\hline
  Precipitate composition of A ($c_A^P$) &  0.04 & 0.037 &0.035 \\
  \hline
  $c_B^P$ &  0.041 & 0.023 & 0.008 \\
  \hline
  $c_C^P$ & 0.919 & 0.94 & 0.957 \\
  \hline\hline
 \end{tabular}\caption{Values of all simulation Variables}\label{sv}
 \end{center}
\end{table}

\section{Computational Algorithm}\label{algo}
\begin{enumerate}
\item Put requisite number of circles of fixed radius (8 and 16 units) at certain inter-particle distance using periodic boundary conditions  in  a $512^2$ matrix so that the area fraction occupied by the circles equals to the intended volume fraction of particles (5\% and 10\%).
\item The mobility of the particles (circles in 2D) is attributed to nearly zero and mobility of other two components are scaled accordingly.
\item Initial composition profile within the particles is taken as the equilibrium composition of C-rich phase from the isothermal section of ternary phase diagram and the composition of A/B components is taken accordingly (refer section~\ref{mm},~\ref{sp}). Now provide a compositional fluctuation of 0.5\% to the composition at each grid point of a $512^2$ matrix.
\item Compute $ g_i $ in real space which is a function of $ c_i \& \chi_{ij} $ where $ i, j = A \& B $
\item Perform forward Fourier transform (using FFTW library~\cite{Frigo}) to convert the real space $  c_i\left(\mathbf{r},t\right)$ and $  g_i\left(\mathbf{r},t\right)$ values to Fourier space values $  \tilde c_i\left(\mathbf{k},t\right)$ and $  \tilde g_i\left(\mathbf{k},t\right)$
\item Get the modified value of   $\tilde c_i\left(\mathbf{k},t+\Delta t\right)$ and $  \tilde g_i\left(\mathbf{k},t+ \delta t\right)$ after calculation of evolution equations for $c_A $ and  $ c_B$ respectively in 2-D Fourier space.
\item Return to real space after performing backward Fourier transform and scale the $ c_i\left(\mathbf{k},t\right)$ values.
\item Iterate steps 4 to 7 again and again upto desired timestep achieved and make sure that the $  \tilde c_i\left(\mathbf{k},t\right)$ and $  \tilde g_i\left(\mathbf{k},t\right)$ values get modified after each timestep.
\item After a certain time interval (timestep$ \times $ stepsize)  save the modified values of $  \tilde c_i\left(\mathbf{k},t\right)$ and $  \tilde g_i\left(\mathbf{k},t\right)$  in a data file.
\item For ternary plotting, we used grey scale representation of gibbs triangle, which is shown in the next page, is a four distinct region, center region darkest, top region  darker, bottom right region dark and bottom left region white. Darker area signifies C-rich region, dark area denotes B-rich region and white area constitutes of A-rich region. Using this approach we are able to distinguish three distinct regions after plotting the modified composition profiles in GNUPLOT.   
\end{enumerate}  
\begin{figure}[h]
\begin{center}
\includegraphics[scale = 0.5]{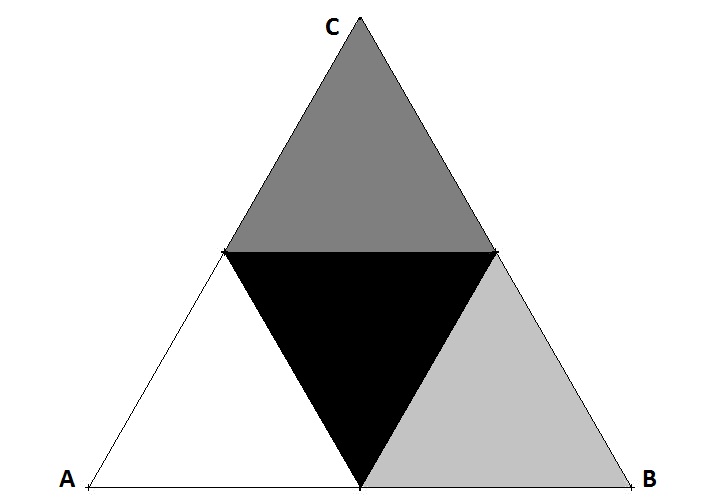}
\caption{Gray scale color map projected on Gibbs triangle}
\label{gs}
\end{center}
\end{figure}
\chapter{Results and Discussion}
In this chapter we present our results obtained after carrying out simulations. The present work is directed at understanding of : 
\begin{itemize}
\item The dynamics of phase separation of a binary mixture ( critical and off-critical compositions) in presence of immobile solid particles. 
\item Influence of preferential wetting on phase evolution.
\item Morphology and domain growth characteristics of such systems.
\item Role of effective interaction energy and relative interfacial energy on the morphology.
\item How particle size, shape and density affect the phase behavior of underlying binary polymeric pattern.
\end{itemize}
 
We have studied the above mentioned aspects following three systems ($S_O, S_W, S_S$) which differs in terms of effective binary interaction parameter ($\chi$), gradient energy parameter ($\kappa$) etc. Each system is simulated for different particle radius and volume fractions. Composition fields of the matrix and particles have considered from the ternary equilibrium phase diagram, described in section ~\ref{tpe}, corresponding to each system. In section ~\ref{iie}, we  show the relative inferfacial energy between the existing phases and produce the equilibrium composition profiles. Microstructures are illustrated in section ~\ref{cmic},~\ref{ocmic}.   
\section{Ternary Phase Equillibria}\label{tpe}
We used three distinct systems (section ~\ref{sp}) for better understanding of our objectives. Three systems of different $\chi$ give rise to three distinct equilibrium phase diagrams. The phase diagrams consist of single phase region (A--rich $\alpha$, B--rich $\beta$, C--rich $\gamma$), two phase region ($\alpha$ +$\beta$, $\beta$ + $\gamma$, $\alpha$ + $\gamma$) and three phase region ($\alpha$ + $\beta$ + $\gamma$). Equilibrium composition of $\alpha$, $\beta$ and $\gamma$ in the three systems are depicted below. In our simulation, composition field inside the particle is taken as equilibrium composition of $\gamma$ phase and the matrix composition is considered as the composition of the point denoted as  '+' in the following ternary isothermal phase diagrams. \\ \\
In case of system--$S_O$ ($\chi_{AB}$ = 2.5, $\chi_{BC}$ = 3.5, $\chi_{AC}$ = 3.5), \\ \\
($c_A^\alpha$, $c_B^\alpha$, $c_C^\alpha$ ) = ( 0.767, 0.174, 0.059 )\\
($c_A^\beta$, $c_B^\beta$, $c_C^\beta$ ) = (0.174, 0.767, 0.059)\\
($c_A^\gamma$, $c_B^\gamma$, $c_C^\gamma$ ) = (0.04, 0.041, 0.919)\\ \\
In case of system--$S_W$ ($\chi_{AB}$ = 2.5, $\chi_{BC}$ = 4.0, $\chi_{AC}$ = 3.5), \\ \\
($c_A^\alpha$, $c_B^\alpha$, $c_C^\alpha$ ) = ( 0.779, 0.168, 0.053)\\
($c_A^\beta$, $c_B^\beta$, $c_C^\beta$ ) = ( 0.159, 0.807, 0.034)\\
($c_A^\gamma$, $c_B^\gamma$, $c_C^\gamma$ ) = (0.037, 0.023, 0.94)\\ \\
In case of system--$S_S$ ($\chi_{AB}$ = 2.5, $\chi_{BC}$ = 5.0, $\chi_{AC}$ = 3.5), \\ \\
($c_A^\alpha$, $c_B^\alpha$, $c_C^\alpha$ ) = ( 0.803, 0.154, 0.043)\\
($c_A^\beta$, $c_B^\beta$, $c_C^\beta$ ) = ( 0.145, 0.843, 0.012)\\
($c_A^\gamma$, $c_B^\gamma$, $c_C^\gamma$ ) = ( 0.035, 0.008, 0.957)\\ \\
\begin{figure}[H]
\begin{center}
\includegraphics[scale=0.4]{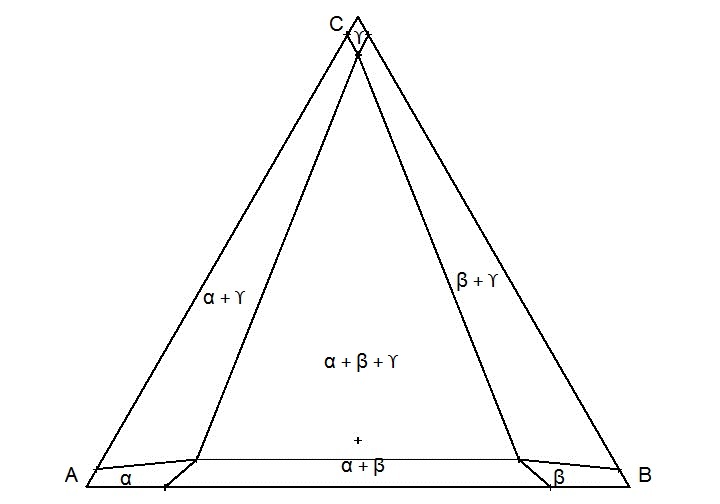}
\caption{Isothermal section of the phase diagram for system $S_O$ ($\chi_{AB}$ = 2.5, $\chi_{BC}$ = 3.5, $\chi_{AC}$ = 3.5)(schematic)}
\label{profile1}
\end{center}
\end{figure}
\begin{figure}[H]
\begin{center}
\includegraphics[scale=0.4]{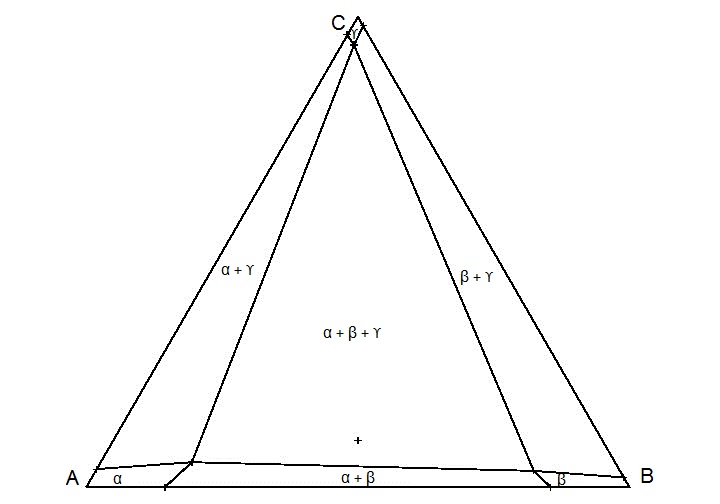}
\caption{Isothermal section of the phase diagram for system $S_W$ ($\chi_{AB}$ = 2.5, $\chi_{BC}$ = 4.0, $\chi_{AC}$ = 3.5)(schematic)}
\label{profile2}
\end{center}
\end{figure}
\begin{figure}[H]
\begin{center}
\includegraphics[scale=0.4]{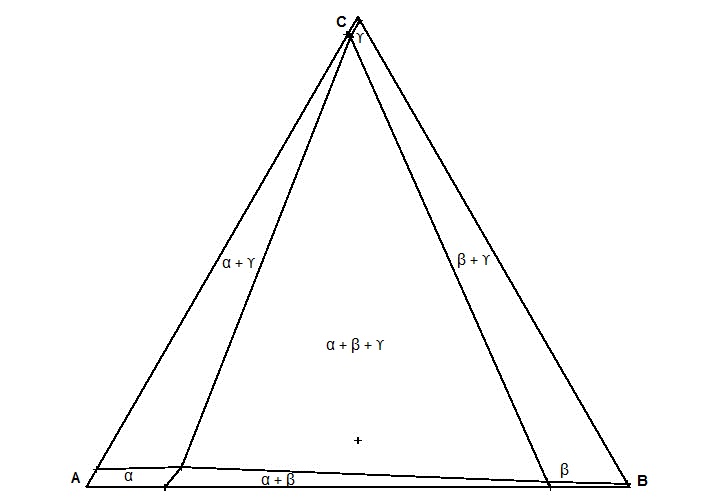}
\caption{Isothermal section of the phase diagram for system $S_S$ ($\chi_{AB}$ = 2.5, $\chi_{BC}$ = 5.0, $\chi_{AC}$ = 3.5)(schematic)}
\label{profile3}
\end{center}
\end{figure}

\section{Interphase Interfacial Energy}\label{iie}
One of our objectives was to study the effect of relative interfacial energies between co-existing phases to the microstructural evolution. For that purpose we have used three different sets of $\kappa_A$, $\kappa_B$, $\kappa_C$ values ( system $S_O$, system - $S_W$, system - $S_S$ respectively). These combinations result in three sets of interfacial energies corresponding to $\alpha - \beta$, $\beta - \gamma$, $\alpha - \gamma$ interfaces.

\begin{figure}[H]
\begin{center}
\includegraphics[scale=0.3]{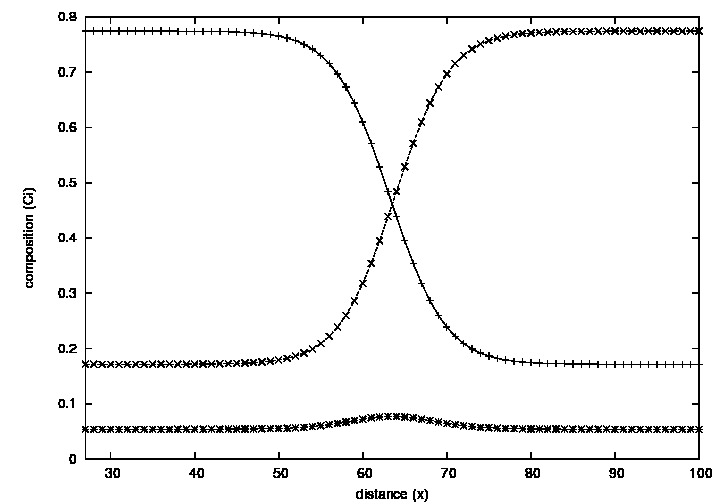}(a)\\
\includegraphics[scale=0.3]{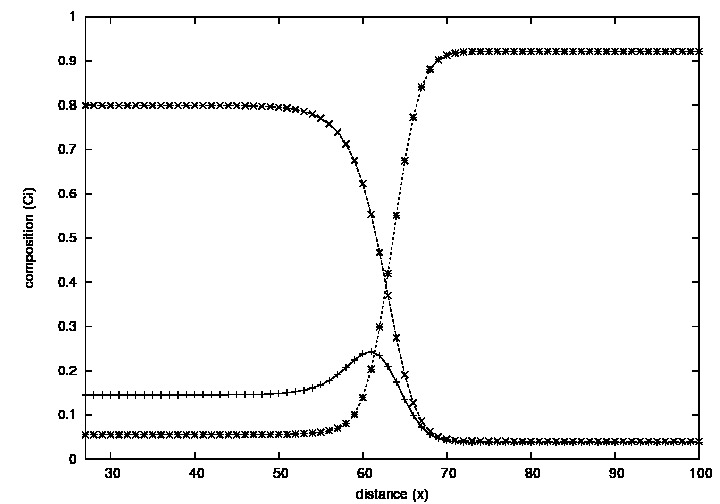}(b)\\
\includegraphics[scale=0.3]{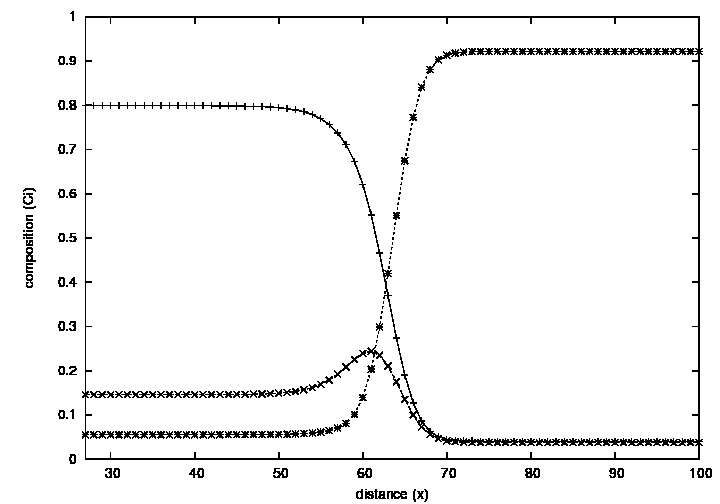}(c)\\
\caption{Equilibrium composition profile across (a) $\alpha - \beta$ interface (b) $\beta - \gamma$ interface (c) $\alpha - \gamma$ interface according to system $S_O$ variables}\label{iep1}
\end{center}
\end{figure}

\begin{figure}[H]
\begin{center}
\includegraphics[scale=0.4]{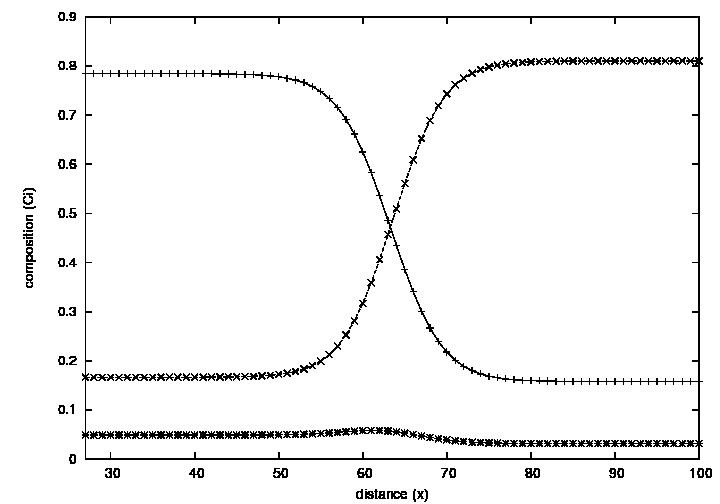}(a)\\
\includegraphics[scale=0.4]{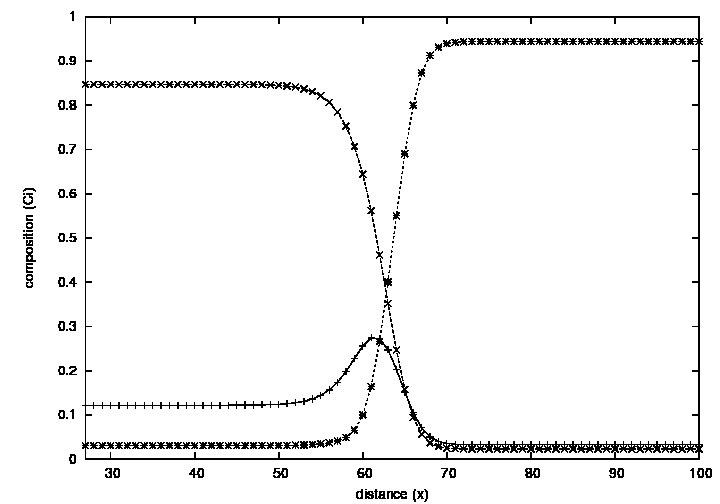}(b)\\
\includegraphics[scale=0.4]{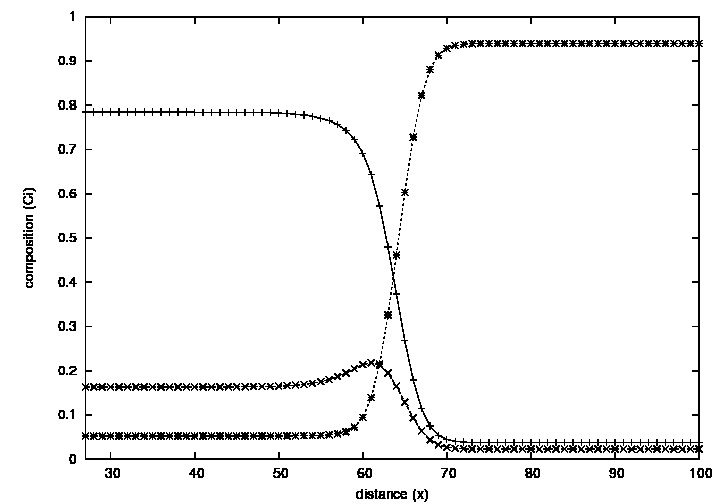}(c)\\
\caption{Equilibrium composition profile across (a) $\alpha - \beta$ interface (b) $\beta - \gamma$ interface (c) $\alpha - \gamma$ interface according to system $S_W$ variables}\label{iep2}
\end{center}
\end{figure}

\begin{figure}[H]
\begin{center}
\includegraphics[scale=0.4]{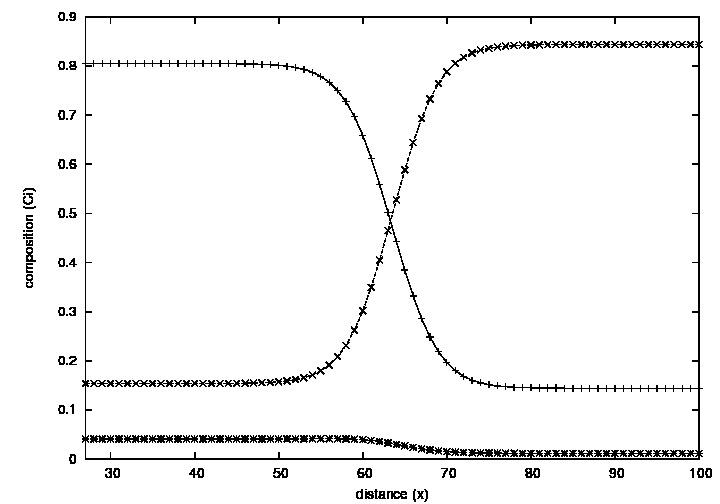}(a)\\
\includegraphics[scale=0.4]{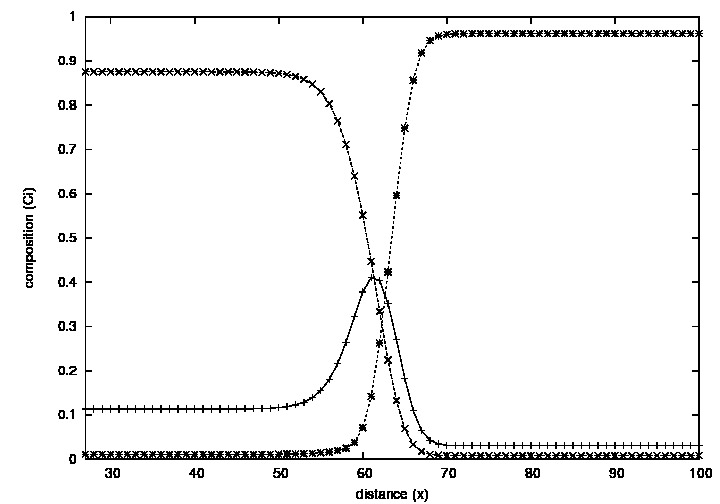}(b)\\
\includegraphics[scale=0.4]{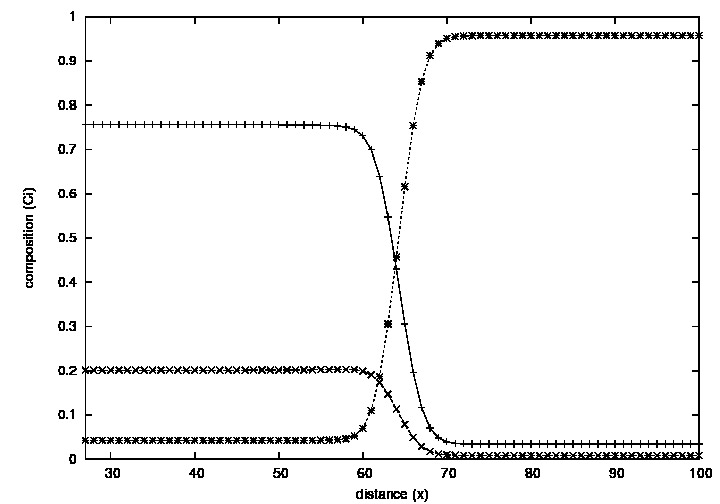}(c)\\
\caption{Equilibrium composition profile across (a) $\alpha - \beta$ interface (b) $\beta - \gamma$ interface (c) $\alpha - \gamma$ interface according to system $S_S$ variables}\label{iep3}
\end{center}
\end{figure}

To calculate interfacial energy of an interface at equilibrium, for example $\alpha - \beta$ interface, a 1-D simulation is set up with one half of the system with $\alpha$ and other half with $\beta$. Then the Cahn-Hilliard equations (eqns. ~\ref{e1}, ~\ref{e2}) are soved to steady state to get the equilibrium composition profiles which are shown in Fig. [\ref{iep1},\ref{iep2},\ref{iep3}]. Relative interfcial energies between co-existing phases, calculated using Eqns. ~\ref{ge1} and ~\ref{ge2}, are listed in Table ~\ref{geie}.
\begin{table}[h]
\begin{center}
\begin{tabular}{|c|c|c|c|c|c|c|}
\hline
system & $\kappa_A$ & $\kappa_B$ & $\kappa_C$ & $\sigma_{\alpha\beta}$ & $\sigma_{\beta\gamma}$ &$\sigma_{\alpha\gamma}$\\
\hline
$S_O$ & 4.0 &4.0 & 4.0 & 0.15 & 0.53 & 0.53 \\
\hline
$S_W$ & 3.0 &5.0 & 5.0 & 0.18 & 0.78 & 0.63 \\
\hline
$S_S$ & 2.0 &6.0 & 6.0 & 0.23 & 1.156  & 0.76  \\
\hline
\end{tabular}\caption{Gradient energy parameters and corresponding interfcial energies}\label{geie}
\end{center}
\end{table}
\newpage

\section{Microstructures in $A_{50}B_{50}$}\label{cmic}
We have considered the matrix to be composed of binary A/B blend in nearly 50 : 50 ratio which signifies to be the critical mixture (composition at the critical point in a temperature vs composition phase diagram). This blend is simulated in presence of randomly distributed stationary spherical particles (C) which occupy volume fractions of 5\% and 10\% respectively. The blend is assumed to be incompressible, that is, the sum of volumes of three phases equal to one. 

All simulations are performed following three systems ($S_O$, $S_W$, $S_S$), tabulated in section~\ref{sp}, which are designed judiciously to incorporate the effects of variable $\chi$, $\kappa$ etc. More precisely, systems are designed in such a way that incorporate no preference ($S_O$), weak preference ($S_W$) and strong preference ($S_S$) for the particles by one of the components (say, A). The ternary microstructures can be interpreted following the gray scale color map [Fig.~\ref{gs}], described in section~\ref{algo}.

\subsection{System $S_O$}
\begin{figure}[H]
\begin{center}
\includegraphics[trim = 50mm 25mm 50mm 25mm, clip, scale=0.5]{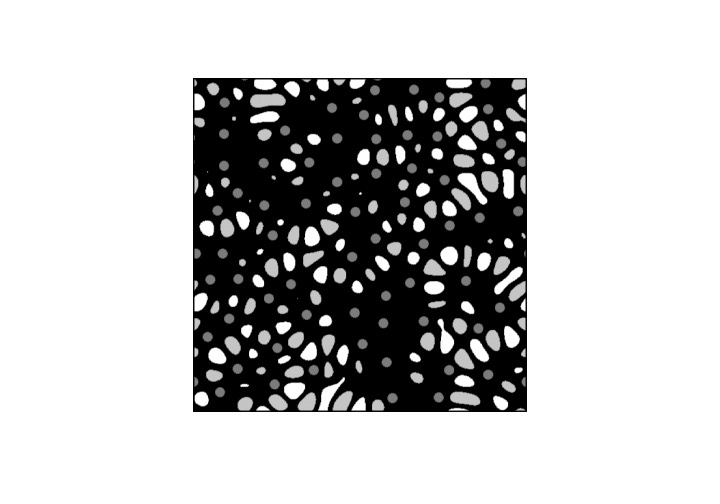}\includegraphics[trim = 50mm 25mm 50mm 25mm, clip,scale=0.5]{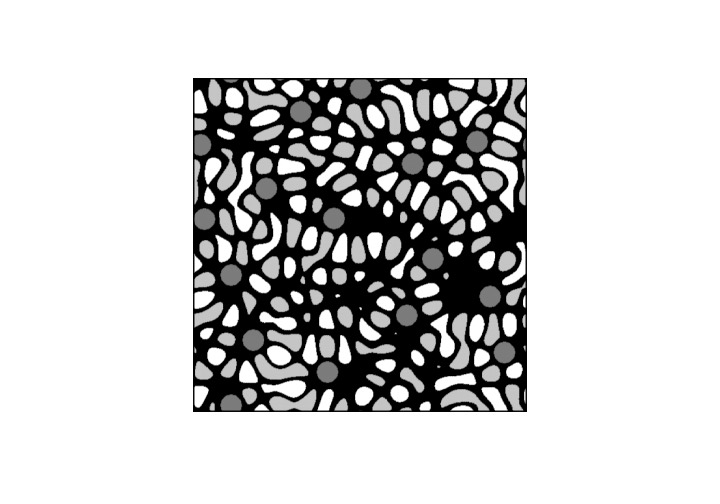}\\
\includegraphics[trim = 50mm 25mm 50mm 25mm, clip, scale=0.5]{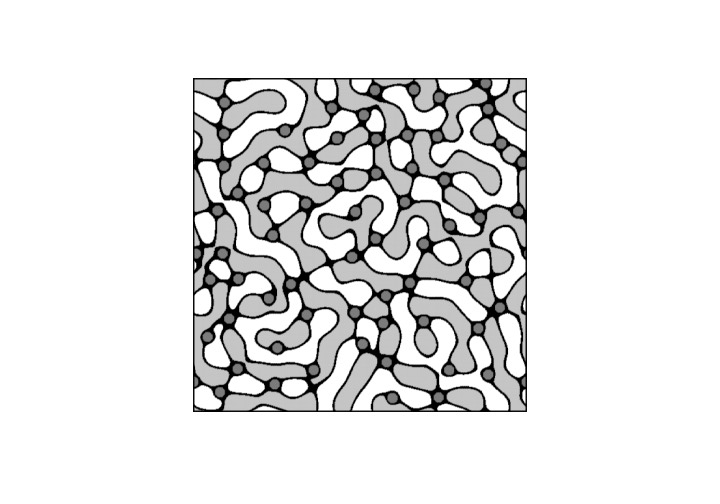}\includegraphics[trim = 50mm 25mm 50mm 25mm, clip,scale=0.5]{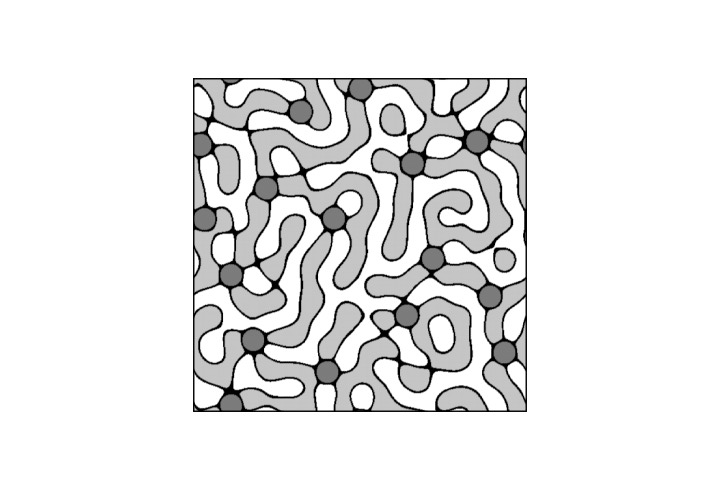}\\
\includegraphics[trim = 50mm 25mm 50mm 25mm, clip, scale=0.5]{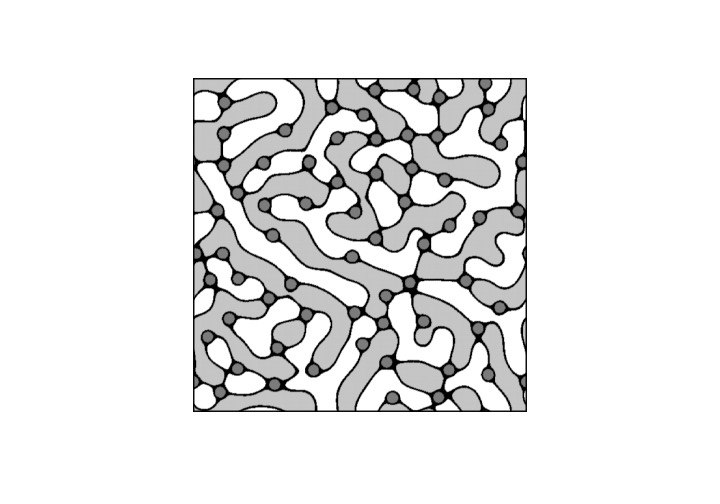}\includegraphics[trim = 50mm 25mm 50mm 25mm, clip,scale=0.5]{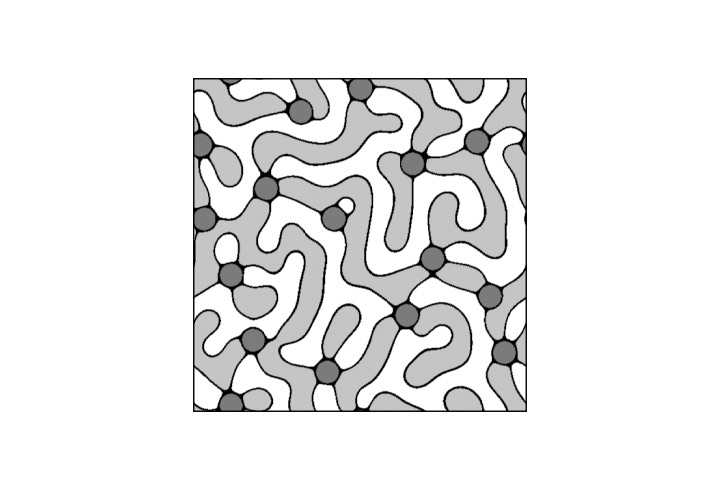}\\
\caption{Microstructures corresponding to left column is for particle radius of 8 units and right column is for particle radius of 16 units for the same volume fraction (\textbf{5\%}) of particles.The top picture is from some early stage (t = 1500 time steps), middle one is of intermediate stage (t = 3000 time steps) and bottom one is for late-stage (t = 5000 timesteps). All corresponding microstructures are compared at similar timestep and follow \textbf{system $S_O$}.}\label{512_p1}
\end{center}
\end{figure}

\begin{figure}[H]
\begin{center}
\includegraphics[trim = 50mm 25mm 50mm 25mm, clip, scale=0.5]{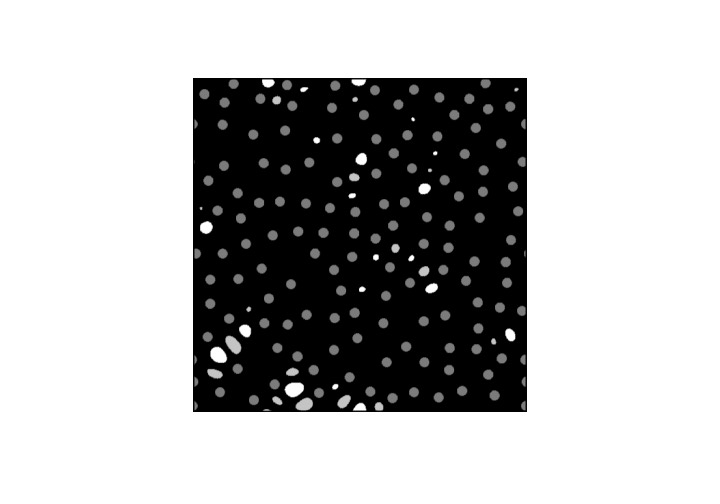}\includegraphics[trim = 50mm 25mm 50mm 25mm, clip,scale=0.5]{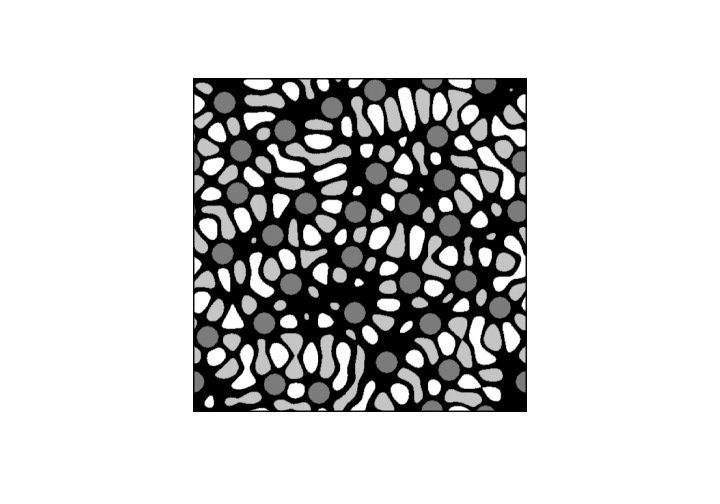}\\
\includegraphics[trim = 50mm 25mm 50mm 25mm, clip, scale=0.5]{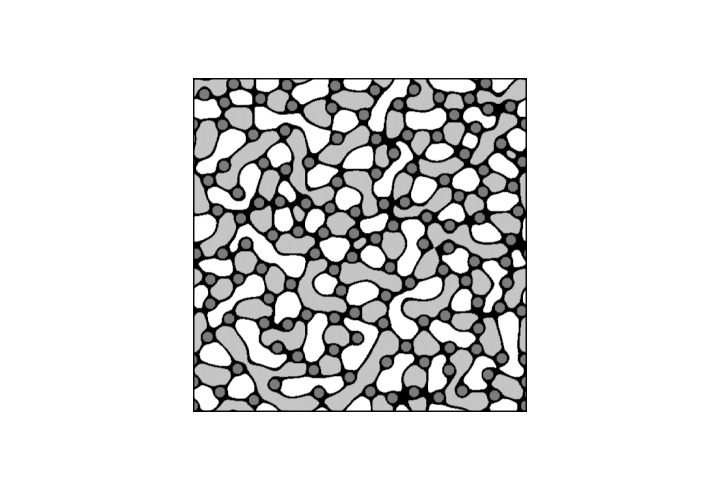}\includegraphics[trim = 50mm 25mm 50mm 25mm, clip,scale=0.5]{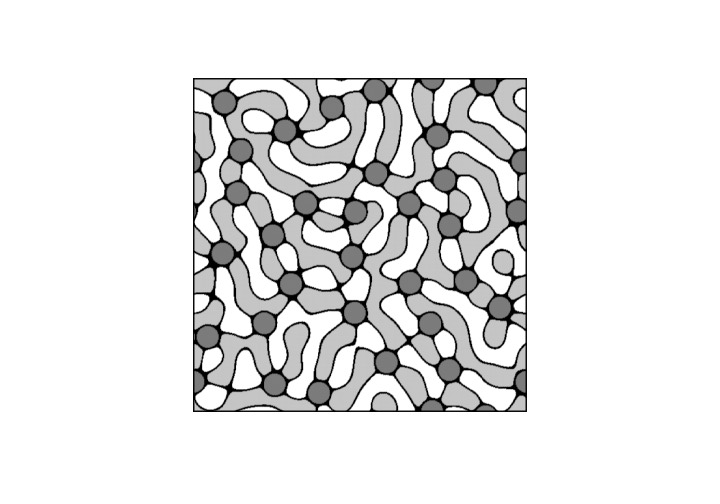}\\
\includegraphics[trim = 50mm 25mm 50mm 25mm, clip, scale=0.5]{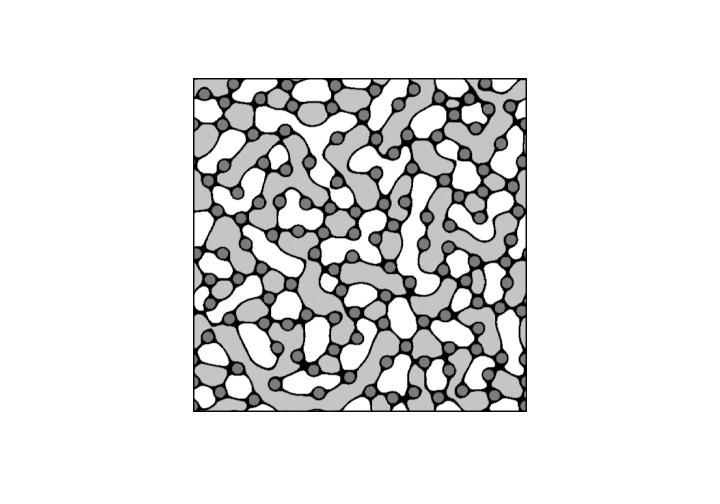}\includegraphics[trim = 50mm 25mm 50mm 25mm, clip,scale=0.5]{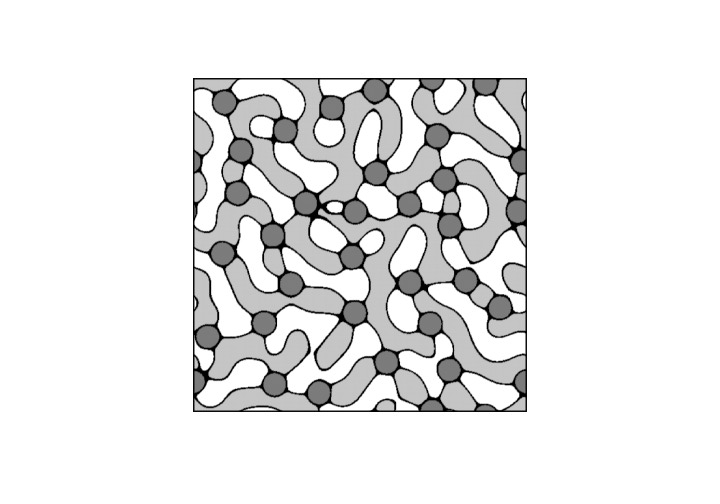}\\
\caption{Microstructures corresponding to left column is for particle radius of 8 units and right column is for particle radius of 16 units for the same volume fraction (\textbf{10\%}) of particles.The top picture is of some early stage (t = 1500 time steps), middle one is of intermediate stage (t = 3000 time steps) and bottom one is of late-stage (t = 5000 timesteps). All corresponding microstructures are compared at similar timestep and follow \textbf{system $S_O$}.}\label{1012_p1}
\end{center}
\end{figure}

 This subsection refers to, when system $S_O$ is attributed to the morphological evolution, for phase separation of a critical binary mixture, in presence of spherical particles. For the case volume fraction of particles is 5\%, the corresponding figure is ~\ref{512_p1}. Figure ~\ref{1012_p1} is the case when volume fraction of the particles is 10\%.  We show three snapshots of each case, one from early stage, one from the intermediate stage and one from last stage. All other necessary details are provided at the caption of each figure. 

According to parameters corresponding to system $S_O$, components do not have any preference for the particles or in other words particles interacting symmetrically to both components. Microstructures corresponding to Fig. [~\ref{512_p1},~\ref{1012_p1}] show that  at early times both phases start appearing simultaneously which eventually leads to nearly complete phase separation at late times and a bi-continuous pattern is observed irrespective of particle size or volume fraction. Interfacial energy driven coarsening takes place after phase separation. Late times morphology demonstrates a surplus of A at some places and surfeit of B at some other places around the particles. Thus both the issues lead to a overall neutral preference of particles for the binary AB.  

Fig. \ref{512_p1} suggests that kinetics of phase separation seems to be slower in presence of smaller particles (keeping volume fraction same). Keeping the particles size same, if volume fraction of particles increases [Fig. ~\ref{1012_p1}] then the kinetics seems to be even more slower. At late times both case shows a bi-continuous pattern. However, domain size is bigger for the case with larger particles, compared to smaller particles. 
\newpage
\subsection{System $S_W$}
\begin{figure}[H]
\begin{center}
\includegraphics[trim = 50mm 25mm 50mm 25mm, clip, scale=0.5]{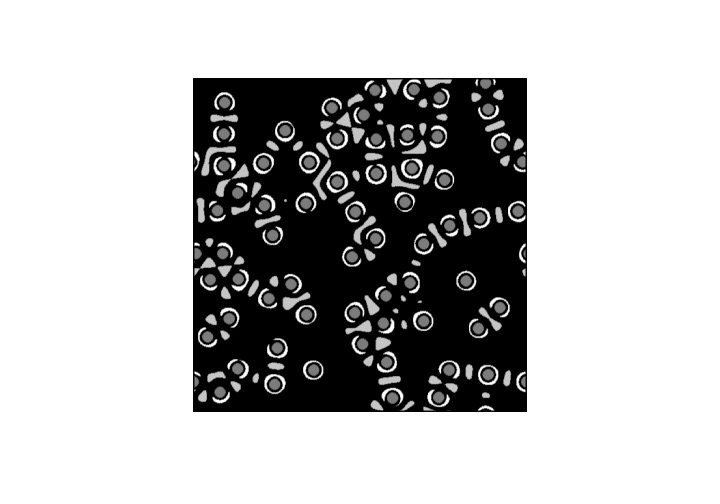}\includegraphics[trim = 50mm 25mm 50mm 25mm, clip,scale=0.5]{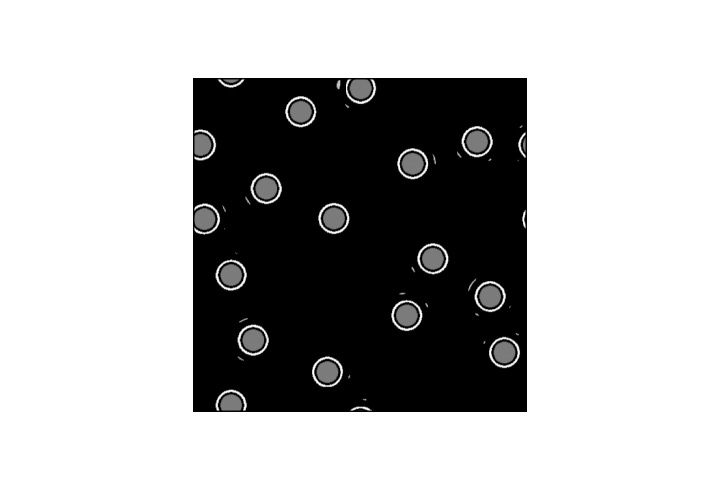}\\
\includegraphics[trim = 50mm 25mm 50mm 25mm, clip, scale=0.5]{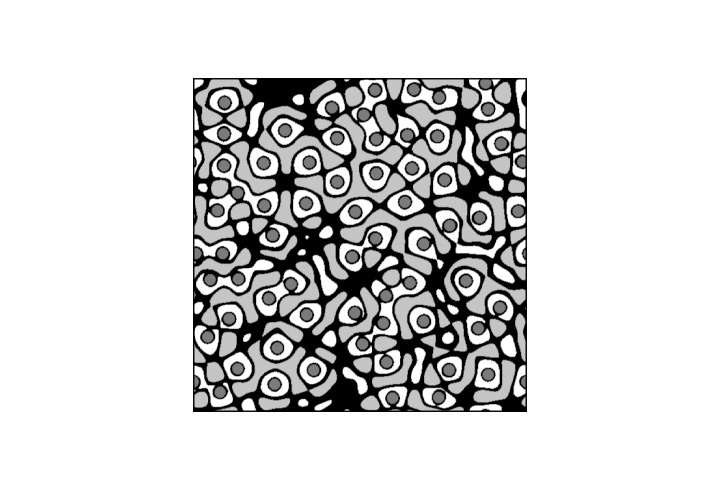}\includegraphics[trim = 50mm 25mm 50mm 25mm, clip,scale=0.5]{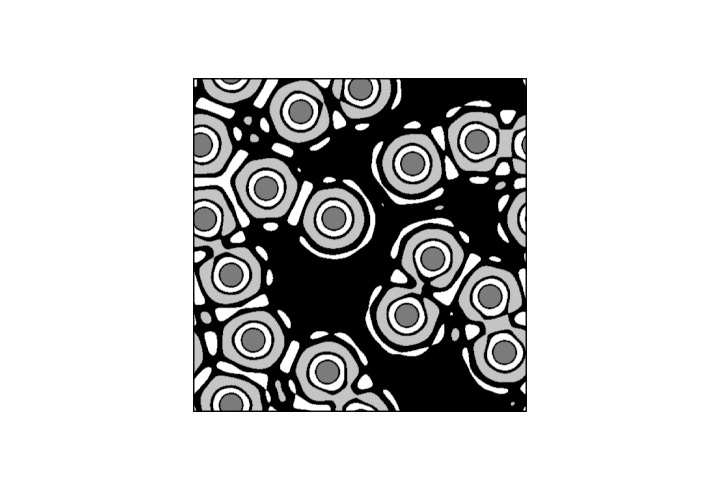}\\
\includegraphics[trim = 50mm 25mm 50mm 25mm, clip, scale=0.5]{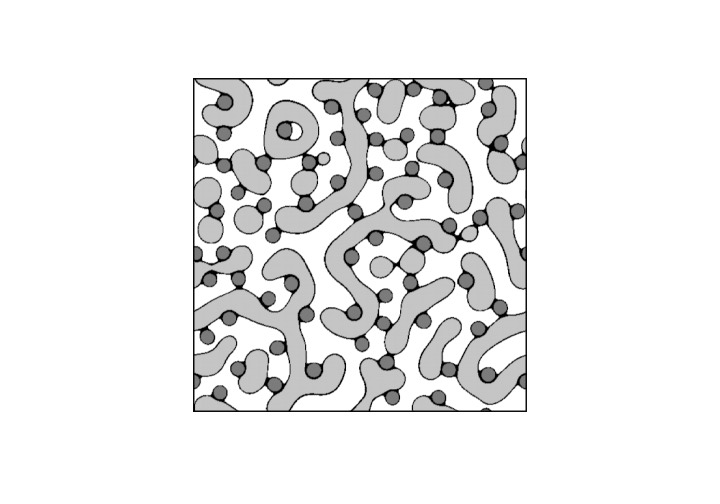}\includegraphics[trim = 50mm 25mm 50mm 25mm, clip,scale=0.5]{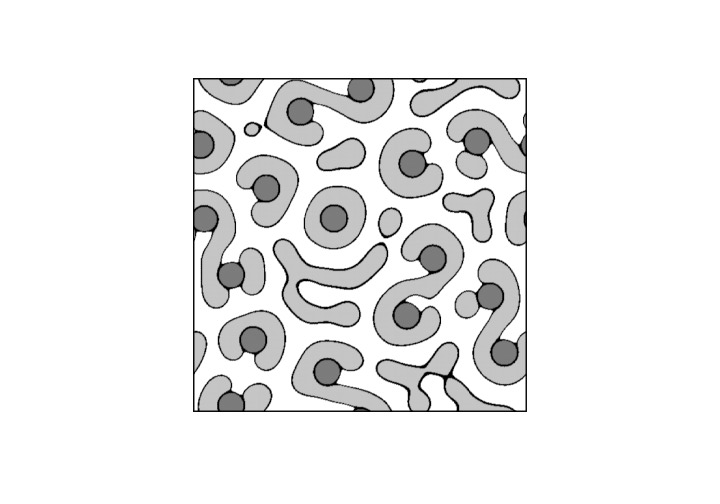}\\
\caption{Microstructures corresponding to left column is for particle radius of 8 units and right column is for particle radius of 16 units for the same volume fraction (\textbf{5\%}) of particles.The top picture is from some early stage (t = 200 time steps), middle one is of intermediate stage (t = 500 time steps) and bottom one is for late-stage (t = 3000 timesteps). All corresponding microstructures are compared at similar timestep and follow \textbf{system $S_W$}.}\label{512_p2}
\end{center}
\end{figure}

\begin{figure}[H]
\begin{center}
\includegraphics[trim = 50mm 25mm 50mm 25mm, clip, scale=0.5]{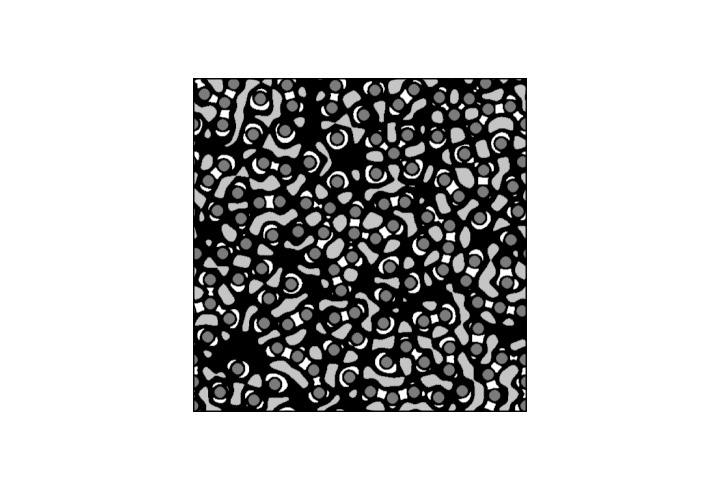}\includegraphics[trim = 50mm 25mm 50mm 25mm, clip,scale=0.5]{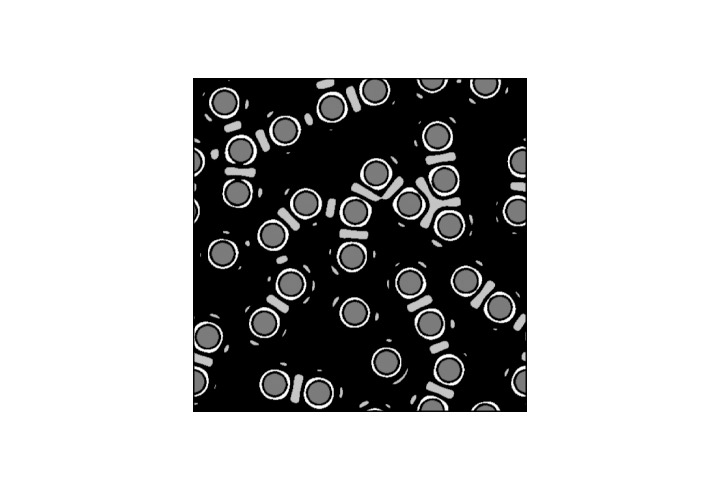}\\
\includegraphics[trim = 50mm 25mm 50mm 25mm, clip, scale=0.5]{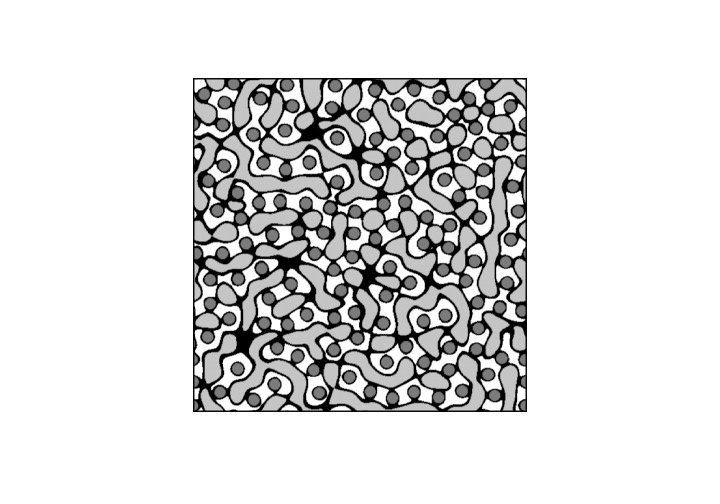}\includegraphics[trim = 50mm 25mm 50mm 25mm, clip,scale=0.5]{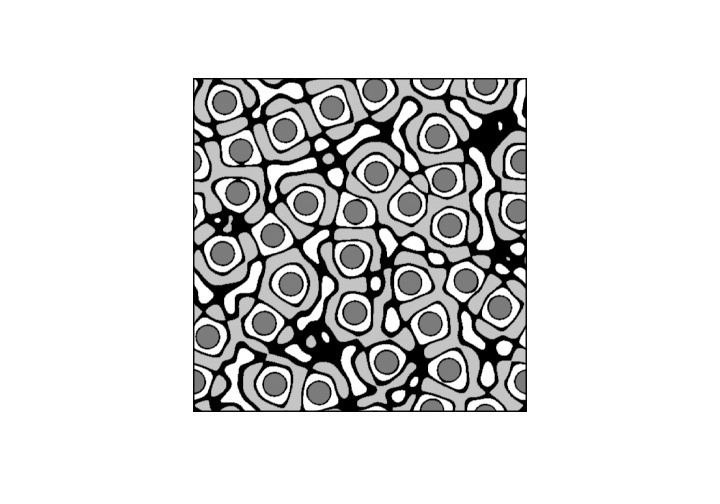}\\
\includegraphics[trim = 50mm 25mm 50mm 25mm, clip, scale=0.5]{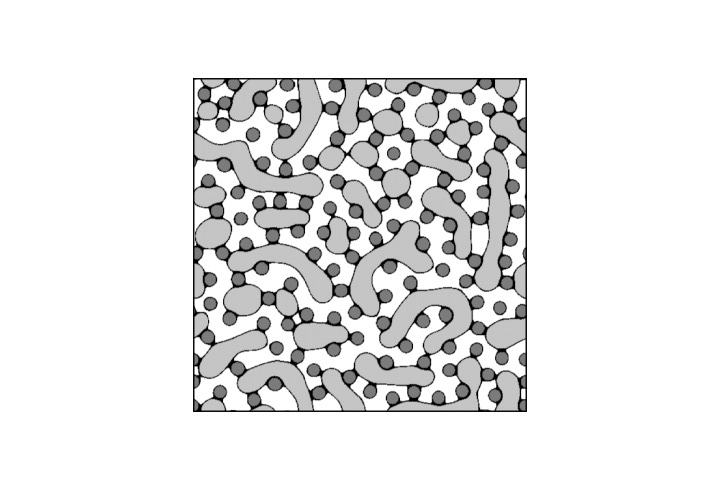}\includegraphics[trim = 50mm 25mm 50mm 25mm, clip,scale=0.5]{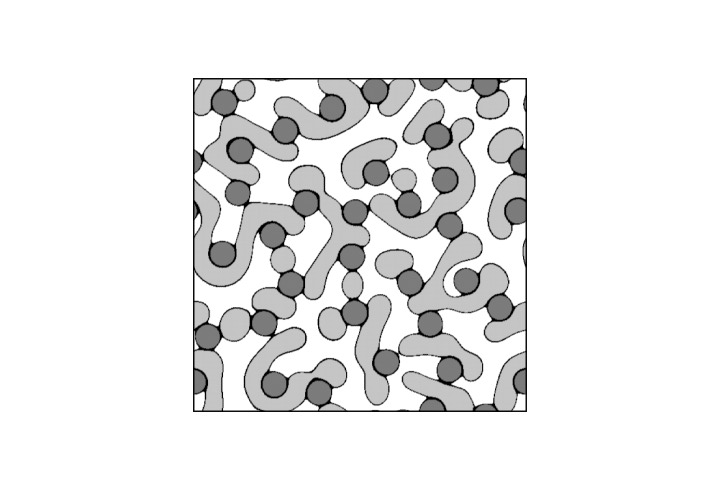}\\
\caption{Microstructures corresponding to left column is for particle radius of 8 units and right column is for particle radius of 16 units for the same volume fraction (\textbf{10\%}) of particles.The top picture is of some early stage (t = 200 time steps), middle one is of intermediate stage (t = 500 time steps) and bottom one is of late-stage (t = 3000 timesteps). All corresponding microstructures are compared at similar timestep and follow \textbf{system $S_W$}.}\label{1012_p2}
\end{center}
\end{figure}

This subsection refers to, when system $S_W$ is attributed to the morphological evolution, for phase separation of a critical binary mixture, in presence of spherical particles. For the case volume fraction of particles is 5\%, the corresponding figure is ~\ref{512_p2}. Figure ~\ref{1012_p2} is the case when volume fraction of the particles is 10\%.  We show three snapshots of each case, one from early stage, one from the intermediate stage and one from last stage. All other necessary details are provided at the caption of each figure.\\
\\
According to the parameters corresponding to system $S_W$, component A is weakly preferred to the particle surface. Early stage to intermediate stage morphology illustrates a ring pattern where particles are completely wetted by A components. This can be referred as core-shell morphology with particle as core and A (preferred component) as shell. As larger size particles are greater distance apart, it seems a thick layer of B forms at the surface of A components. However, for smaller size particles (no. of particles are more to keep the volume fraction same) the process of engulfing of A around the particles breaks the  nucleating domains of B. Hence, for such case the concentric circular layer of B is not apparent. Moreover, Fig. [~\ref{512_p2},~\ref{1012_p2}] clearly demonstrate that phase separation is not complete at late times, and bi-continuous morphology is also not observed. The late stage morphology consists of isolated B domains in a continuous sea of A. Coarsening prevails to these domains to reduce the interfacial energy. It is obvious from the late stage behavior, that the phase evolution and growth in A/B binary happens in such a fashion that particles appear at the interfaces. It is worthwhile to mention that for all the cases early stage core-shell morphology breaks towards late stage, although in few cases (\ref{512_p2}, R = 16 units) such morphology remains but with a different shell, i.e., B, around the core particles.  
\newpage
\subsection{System $S_S$}
\begin{figure}[H]
\begin{center}
\includegraphics[trim = 50mm 25mm 50mm 25mm, clip, scale=0.5]{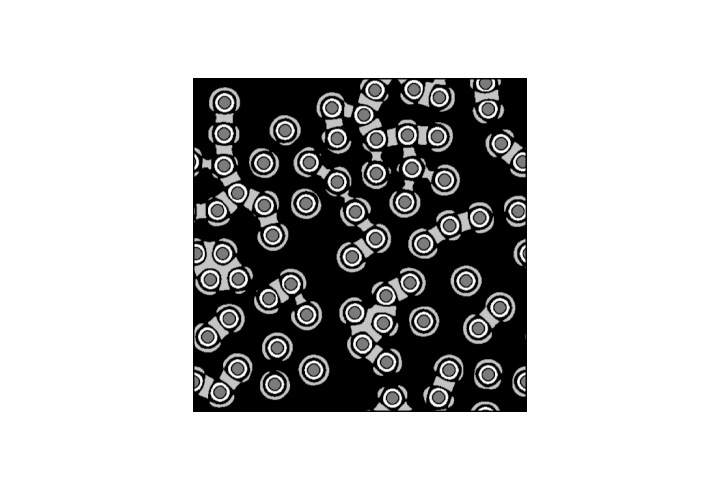}\includegraphics[trim = 50mm 25mm 50mm 25mm, clip,scale=0.5]{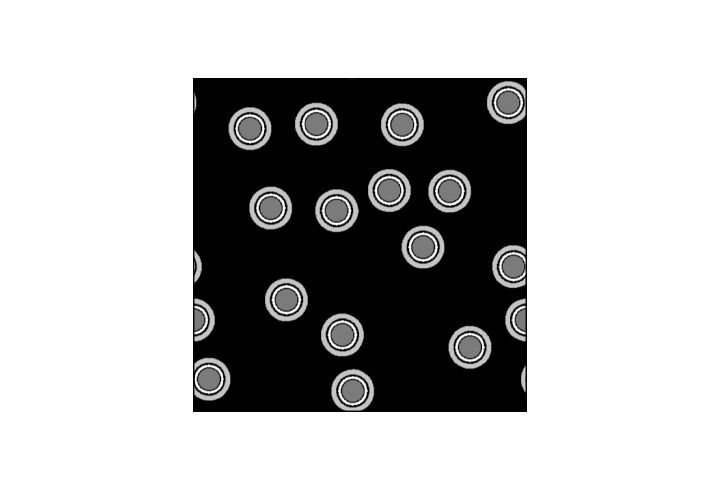}\\
\includegraphics[trim = 50mm 25mm 50mm 25mm, clip, scale=0.5]{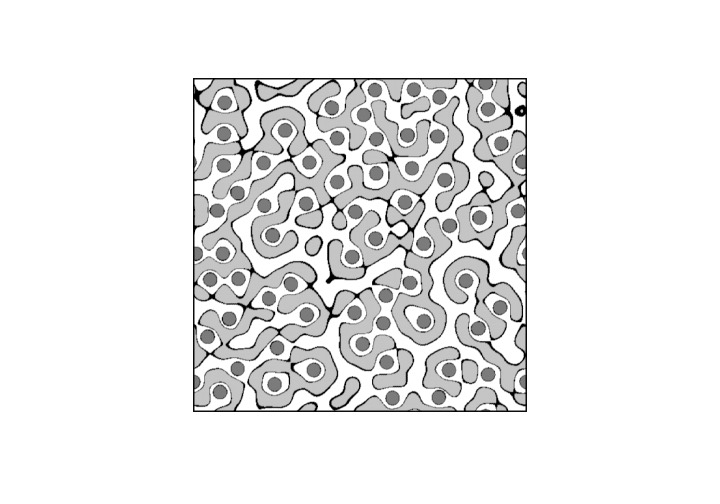}\includegraphics[trim = 50mm 25mm 50mm 25mm, clip,scale=0.5]{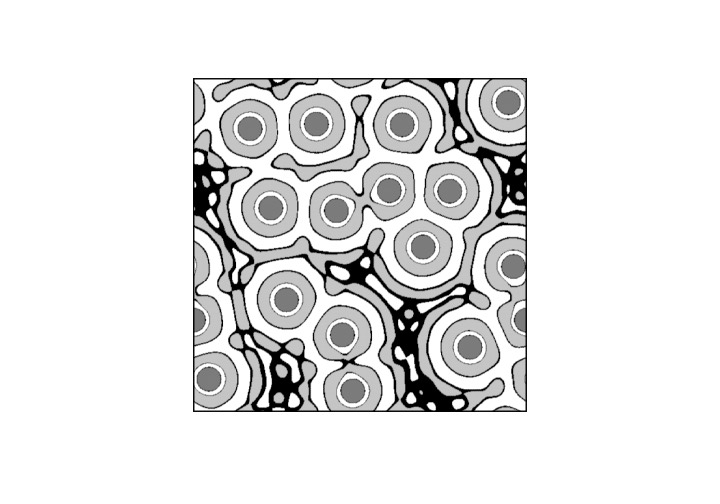}\\
\includegraphics[trim = 50mm 25mm 50mm 25mm, clip, scale=0.5]{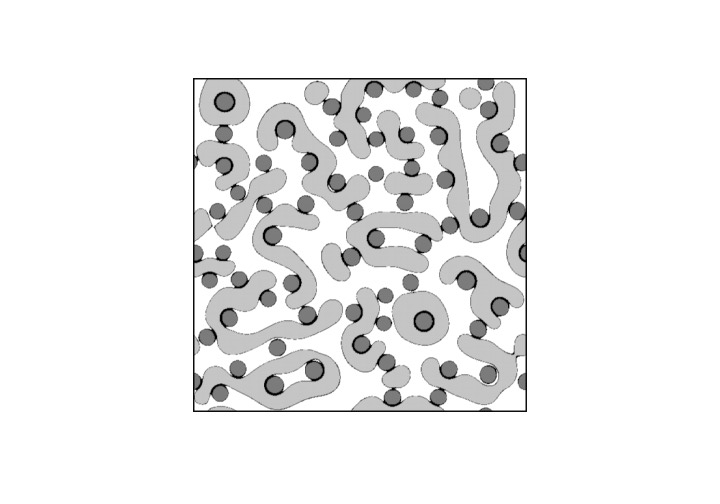}\includegraphics[trim = 50mm 25mm 50mm 25mm, clip,scale=0.5]{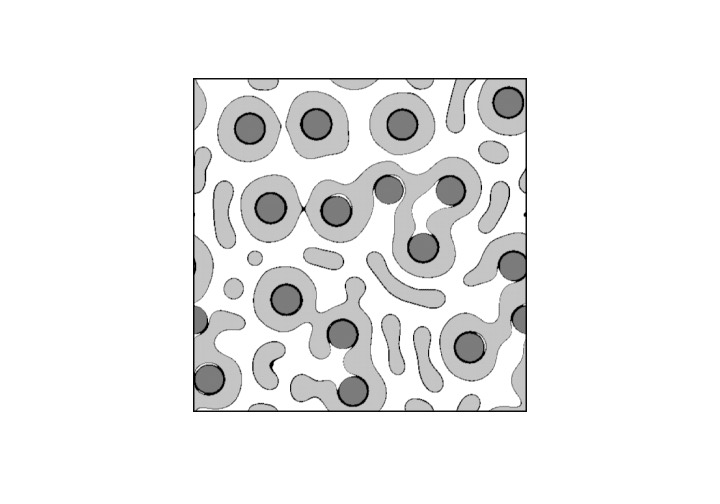}\\
\caption{Microstructures corresponding to left column is for particle radius of 8 units and right column is for particle radius of 16 units for the same volume fraction (\textbf{5\%}) of particles.The top picture is from some early stage (t = 50 time steps), middle one is of intermediate stage (t = 500 time steps) and bottom one is for late-stage (t = 3000 timesteps). All corresponding microstructures are compared at similar timestep and follow \textbf{system $S_S$}.}\label{512_p3}
\end{center}
\end{figure}

\begin{figure}[H]
\begin{center}
\includegraphics[trim = 50mm 25mm 50mm 25mm, clip, scale=0.5]{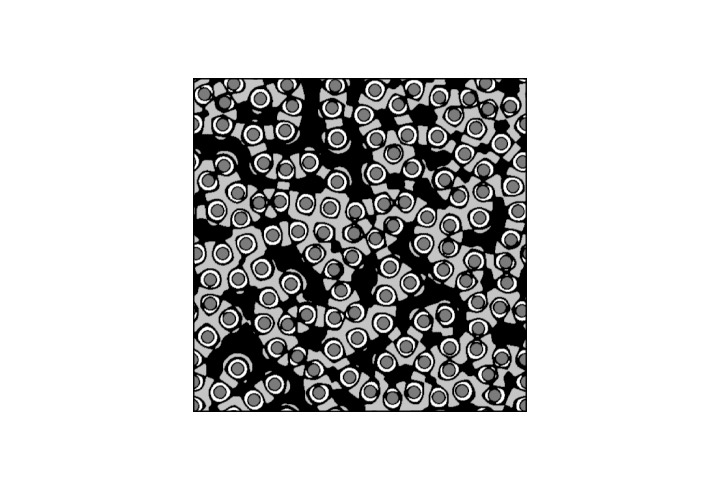}\includegraphics[trim = 50mm 25mm 50mm 25mm, clip,scale=0.5]{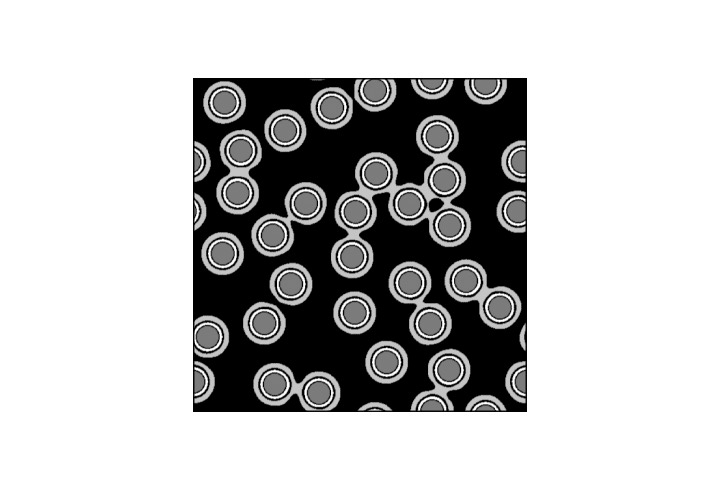}\\
\includegraphics[trim = 50mm 25mm 50mm 25mm, clip, scale=0.5]{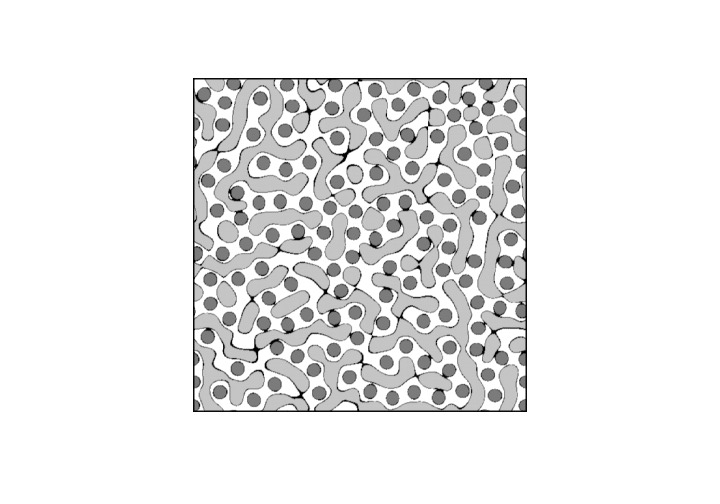}\includegraphics[trim = 50mm 25mm 50mm 25mm, clip,scale=0.5]{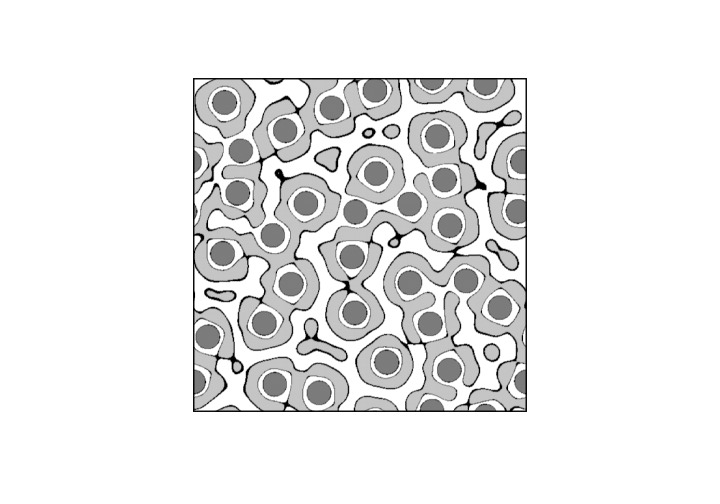}\\
\includegraphics[trim = 50mm 25mm 50mm 25mm, clip, scale=0.5]{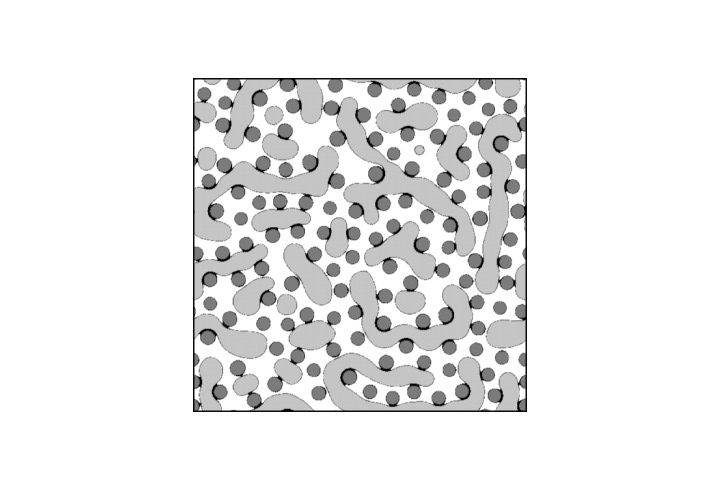}\includegraphics[trim = 50mm 25mm 50mm 25mm, clip,scale=0.5]{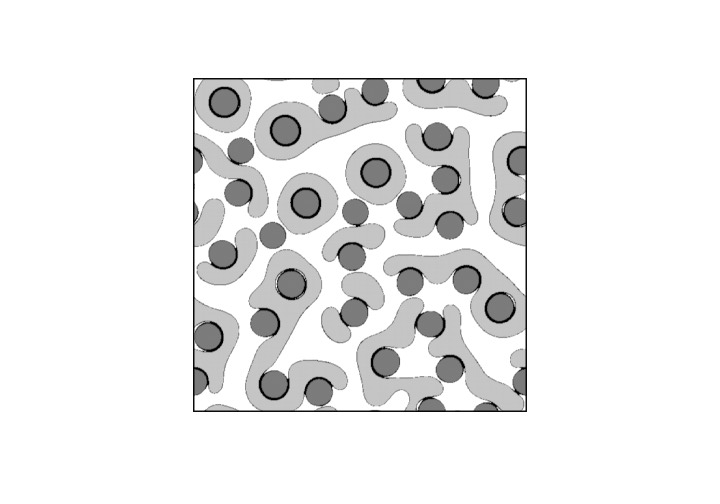}\\
\caption{Microstructures corresponding to left column is for particle radius of 8 units and right column is for particle radius of 16 units for the same volume fraction (\textbf{10\%}) of particles.The top picture is of some early stage (t = 50 time steps), middle one is of intermediate stage (t = 500 time steps) and bottom one is of late-stage (t = 3000 timesteps). All corresponding microstructures are compared at similar timestep and follow \textbf{system $S_S$}.}\label{1012_p3}
\end{center}
\end{figure}

This subsection refers to, when system $S_S$ is attributed to the morphological evolution, for phase separation of a critical binary mixture, in presence of spherical particles. For the case volume fraction of particles is 5\%, the corresponding figure is ~\ref{512_p3}. Figure ~\ref{1012_p3} is the case when volume fraction of the particles is 10\%.  We show three snapshots of each case, one from early stage, one from the intermediate stage and one from last stage. All other necessary details are provided at the caption of each figure.

According to the parameters corresponding to system $S_S$, component A is strongly preferred to the particle surface. At early times, microstructure consists of concentric rings of A and B around the particles. Such core-shell morphology survives longer due to higher interactions between A and particles. Late times, interfacial energy driven coarsening dominates and particles tend to arrange at the interfaces between co-existing phases. Furthermore, the morphology at late times almost resembles to that of system $S_W$, where B components are distributed as isolated islands in a continuous A phase, and characteristic domain size of B is smaller in case of systems with higher particle density. In addition, here also early stage core-shell morphology with shell A breaks towards late stage, forming a different shell, i.e., B, around the core particles.
\newpage

\subsection{Discussion}\label{cdiss}
This subsection is an account to address the microstructures obtained when a binary mixture is quenched critically ($A_{50}B_{50}$). Here, we systematically study the effect of spherical solid particles on the phase evolution due to background spinodal decomposition of a critical binary mixture. When the particles are neutral to A and B (system $S_O$) then ring pattern does not form at early times. Whereas, if the particles are preferably wetted by one of the components (system $S_W/S_S$), ring pattern appears in the microstructure. The ring or alternate layers of A and B forms due to the propagation of composition waves originated from the particle surface~\cite{Lee}. Preferential interaction between component A and particle develops a higher concentration of A near the particle surface than in the bulk. So, there must be depletion of A somewhere, which is immediate next to the A enriched layer away from particle surface. This explains the formation of alternate layers of preferred and non-preferred components around the particle. In case of system $S_O$, the surface directed composition waves are composed of equally likely phases A/B. Hence, ring morphology does not form and A/B domains appear irregularly in the matrix. 
\\ \\
The early to intermediate stage morphology in systems $S_W$ and $S_O$ can be referred as core-shell morphology. As component A is thermodynamically preferable to the particles, A forms the shell with particles at core. At late times few core-shell structure still remains, but now B (non-preferred component) serves as shell around particles. Such shell structure transition can be due to Gibbs Thomson effect~\cite{Porter}. Large radius of curvature of the A layer around the particles causes a higher concentration gradient of A in the shell relative to the matrix. Such concentration gradient expels A from the particle surface, allowing the B to form shell about particles.    
\\ \\
During intermediate to late times the particle density at the interface increases. Consider the two case: (a) a particle is completely enveloped by a preferred component (say, A) and (b) it resides at the interface between co-existing phases (A/B). Second case is more concerned in terms of lowering the interfacial energy of the system. Roughly speaking, when a particle stays in interface than in bulk, the system reduces its energy by an amount $\Pi\gamma_{AB}R_N^2$, where $R_N$ is the particle radius and $\gamma_{AB}$ is the A/B interfacial energy~\cite{Hore}. At early stages the particles may be covered by the favorable components due to the constraints of energetic parameters imposed on them. But, at late times particle density increases at the interface to accommodate the domain coarsening and hence, reduce the total interfacial energy.
\\ \\
 We can explain the increase of particle density at the interface by the following arguments. Morphology of a polymer blend depends on interfacial tension and as well as polarity (comparable to $ \chi $) between the components~\cite{Benderly}. In case of system $S_W/S_S$, the values of energetic parameters ($\chi$ and $\kappa$) confirmed  $\gamma_{BC} > \gamma_{AC}$ and hence A is preferentially attracted to the C (particle) interface. However, at late times particles tend to sit at the interface, though there is a surplus of preferred component A at the interface. This mimics a situation like morphology transition from complete wetting (CW) to partial wetting (PW). Now, to achieve such a partially wet morphology, one of the following conditions has to be satisfied.
\begin{equation}\label{young} \gamma_{AB} > |\gamma_{BC} - \gamma_{AC}| \end{equation}
\begin{equation}\label{ckr}\chi_{AB} > |\chi_{BC} - \chi_{AC}|\end{equation}
 If free energy can be described by a regular solution model, gradient energy coefficient is directly related to pairwise interaction parameter~\cite{Cahn} by the relation $\kappa = \chi a^2/2$, where a is the interaction distance. So, gradient energy and interaction energy both affects the wetting behavior of the particles.Therefore, if both conditions [~\ref{young},~\ref{ckr}] are satisfied by imposed parameters, complete wetting prevails. 
\\ \\
Another explanation related to the interfacial segregation of particles is found in literarure~\cite{Benderly}. Spreading co-efficient (S) is defined as $ S = \gamma_{BC} - \gamma_{AC} - \gamma_{AB} $. S > 0 yields CW morphology and S < 0 results into PW morphology. This parameter stems from the young equation~\ref{young} and equally useful to predict the thermodynamically preferred morphology. However, kinetic factors can prevent us to achieve that morphology. Viscosity, quench depth ($ (\chi - \chi_c)/\chi_c = (T - T_c)/T_c $) etc. are well-known kinetic factors. In our case for system $S_W/S_S$, the thermodynamically favored morphology is CW. However, we obtain a PW morphology. Higher viscosity and deeper quench depth may be responsible for this. Viscosity is dependent on composition. As the morphology approaches to co-continuous, then the interlocking tendency between the co-existing phases increases. Therefore the blend viscosity or resistance to flow increases. Deeper quench depth (system $S_W/S_S$) also seems to develop this type of interlocking and encapsulation is impeded at late times.   
\\ \\
In case of system $S_W/S_S$, the late time morphology demonstrates a single infinite domain of the wetting phase (A) and the B components are trapped as isolated islands in this domain~\cite{Lee}. The process of engulfing of the preferred component around the particle and subsequent domain coarsening effectively breaks the continuity of domain B. Further coarsening of B domains is inhibited as the particles behave as obstacles to the motion of interfaces. It is obvious from late stage microstructures that bi-continuous morphology does not form and hence, phase separation is also incomplete. To elucidate the effect of wetting on the slowing down of domain growth, we performed one set of simulation where components have no preference for the filler (system $S_O$). At late times, this results into complete phase separation , assured by a bi-continuous pattern. Thus, effect of wetting on the slowing down of domain growth is confirmed.
\\ \\
Particle size,shape and density also affects the dynamics of phase separation of a critical binary mixture. Comparing the microstructures obtained, it is qualitatively true that higher volume fraction and smaller radius of particles are more effective for slowing down the kinetics of phase separation and domain growth. Homogeneous distribution of particles leads to effective increase in the viscosity. As smaller particles are distributed more homogeneously compared to bigger particles, viscosity increment is pronounced in case of smaller particles. Moreover, volume fraction is also directly related with the viscosity~\cite{Hore}. There are two competing factors at late times: a) temporal decrease in interfacial tension and b) increase in viscosity. For smaller particles second effect is more striking than the first one and slower kinetics is more pronounced. In case of bigger particles factor (a) dominates the other and hence, kinetics of phase separation and domain growth is faster compared to smaller particles.       
\newpage
\enlargethispage{1in}
\section{Off-symmetric alloys in system $S_S$}\label{ocmic}
This section is meant for presenting the microstructures obtained when a binary alloy of off-critical composition undergoes phase separation in presence of spherical particles. Particle characteristics are similar to the same used in case of binary critical mixture. Our objective was to probe the microstructures in the following two cases:
\begin{itemize}
\item minor component wets the particle surface 
\item major component wets the particle surface
\end{itemize} 
That's why we simulated microstructures for two alloys $A_{40}B_{60}$ and $A_{60}B_{40}$. As system $S_S$ is best suited to fulfill our purpose, we only run the simulation for the case of system $S_S$ parameters enforced on the morphological evolution of off-critical binary alloys. 
\subsection{$A_{40}B_{60}$}
\begin{figure}[H]
\begin{center}
\includegraphics[trim = 50mm 25mm 50mm 25mm, clip, scale=0.5]{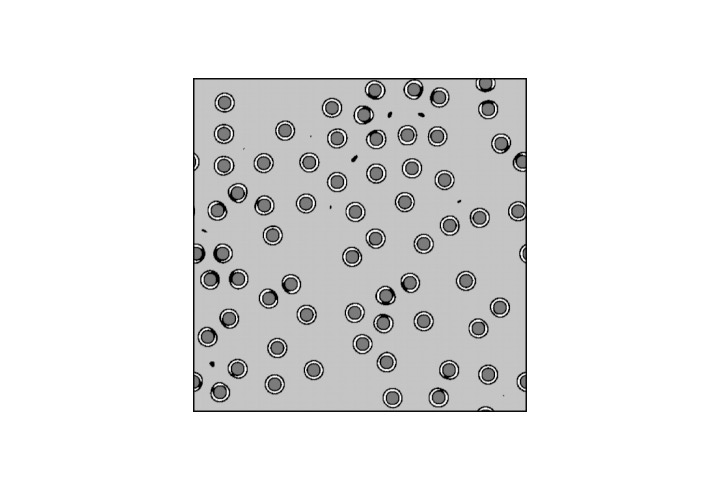}\includegraphics[trim = 50mm 25mm 50mm 25mm, clip,scale=0.5]{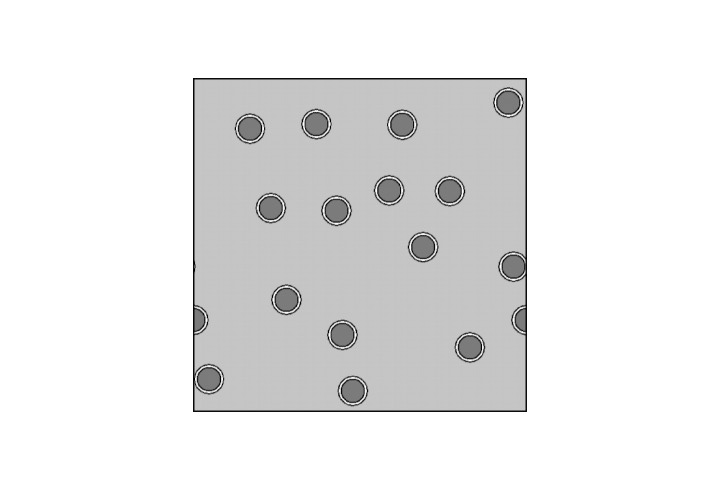}\\
\includegraphics[trim = 50mm 25mm 50mm 25mm, clip, scale=0.5]{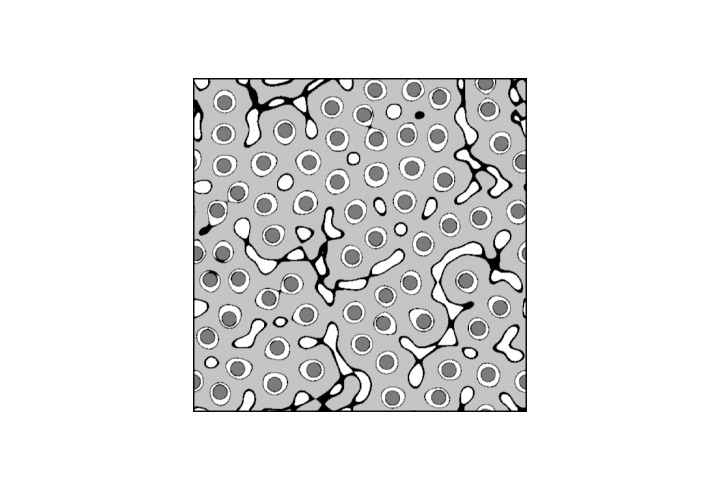}\includegraphics[trim = 50mm 25mm 50mm 25mm, clip,scale=0.5]{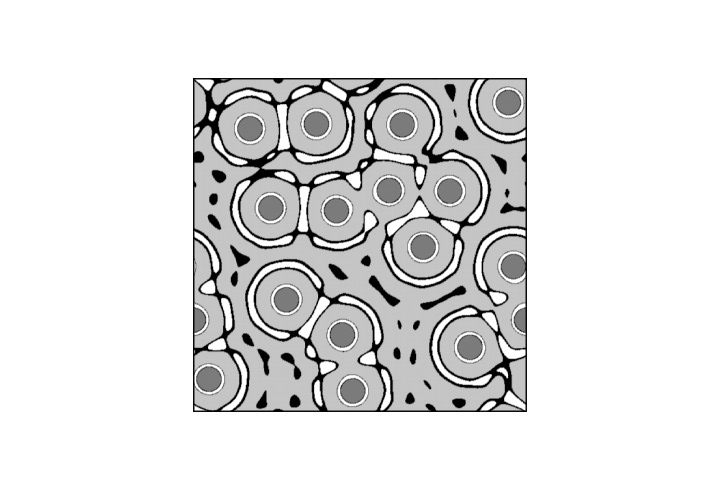}\\
\includegraphics[trim = 50mm 25mm 50mm 25mm, clip, scale=0.5]{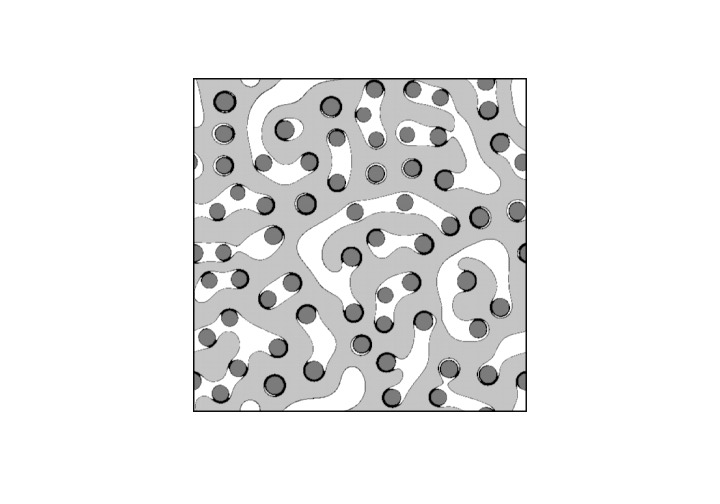}\includegraphics[trim = 50mm 25mm 50mm 25mm, clip,scale=0.5]{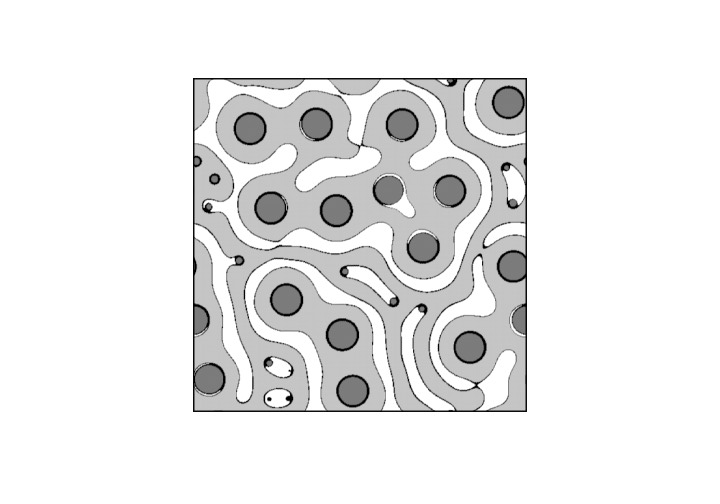}\\
\caption{Microstructures($A_{40}B_{60}$) corresponding to left column is for particle radius of 8 units and right column is for particle radius of 16 units for the same volume fraction (\textbf{5\%}) of particles.The top picture is from some early stage (t = 100 time steps), middle one is of intermediate stage (t = 500 time steps) and bottom one is for late-stage (t = 3000 timesteps). All corresponding microstructures are compared at similar timestep and follow \textbf{system $S_S$}.}\label{o1512_p3}
\end{center}
\end{figure}

\begin{figure}[H]
\begin{center}
\includegraphics[trim = 50mm 25mm 50mm 25mm, clip, scale=0.5]{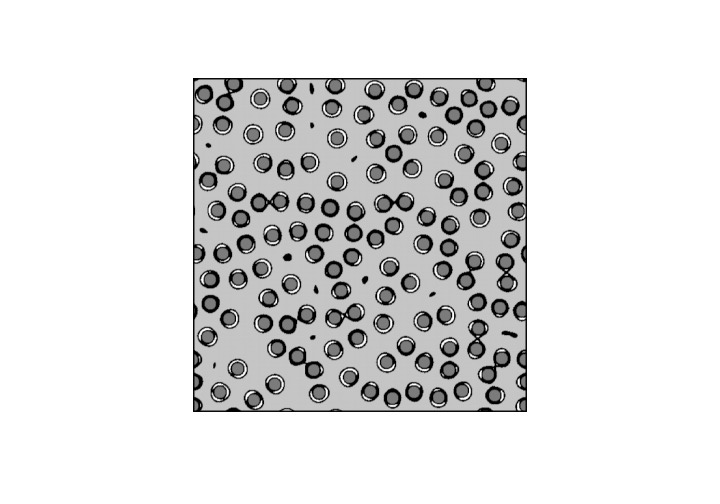}\includegraphics[trim = 50mm 25mm 50mm 25mm, clip,scale=0.5]{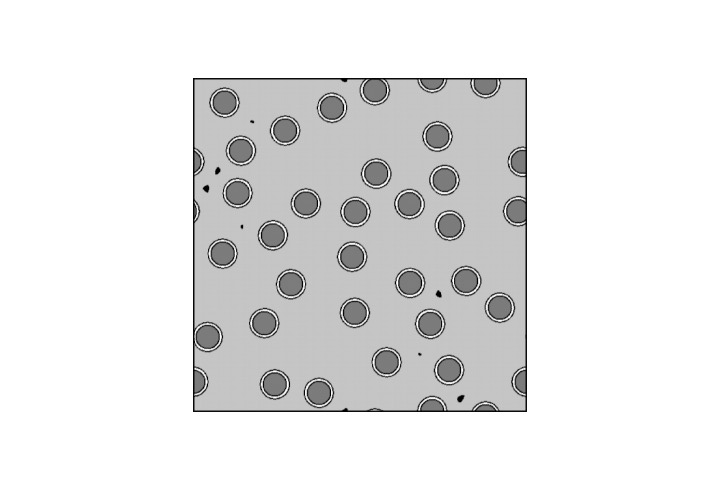}\\
\includegraphics[trim = 50mm 25mm 50mm 25mm, clip, scale=0.5]{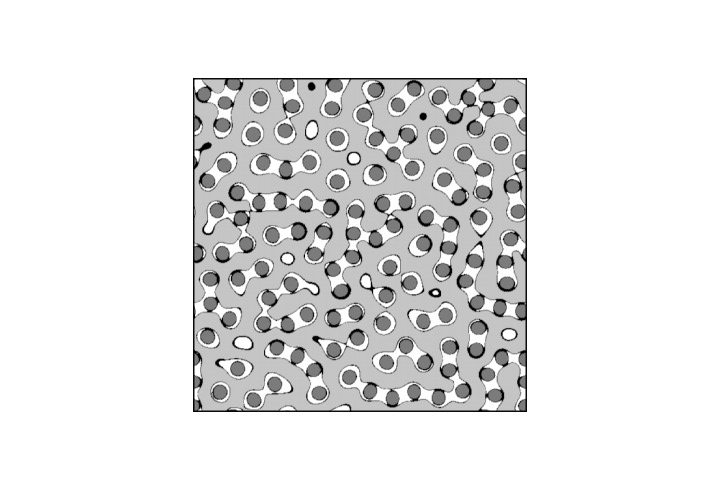}\includegraphics[trim = 50mm 25mm 50mm 25mm, clip,scale=0.5]{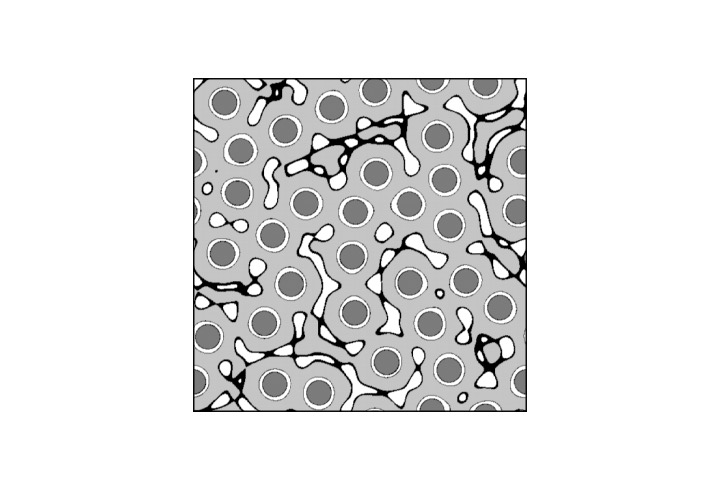}\\
\includegraphics[trim = 50mm 25mm 50mm 25mm, clip, scale=0.5]{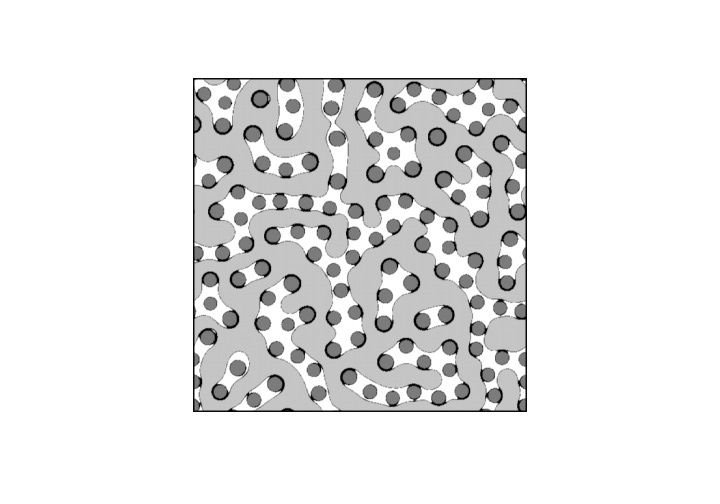}\includegraphics[trim = 50mm 25mm 50mm 25mm, clip,scale=0.5]{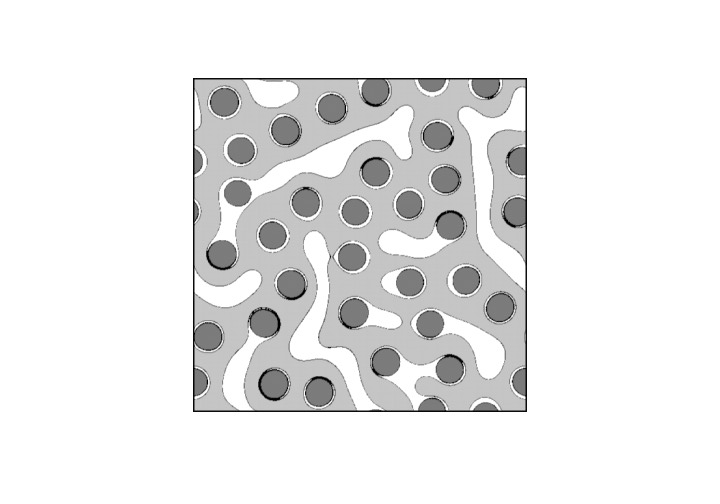}\\
\caption{Microstructures ($A_{40}B_{60}$) corresponding to left column is for particle radius of 8 units and right column is for particle radius of 16 units for the same volume fraction (\textbf{10\%}) of particles.The top picture is of some early stage (t = 100 time steps), middle one is of intermediate stage (t = 500 time steps) and bottom one is of late-stage (t = 3000 timesteps). All corresponding microstructures are compared at similar timestep and follow \textbf{system $S_S$}.}\label{o11012_p3}
\end{center}
\end{figure}

 In this case minor component (A) wets the particle. For the case volume fraction of particles is 5\%, the corresponding figure is ~\ref{o1512_p3}. Figure ~\ref{o11012_p3} is the case when volume fraction of the particles is 10\%.  We show three snapshots of each case, one from early stage, one from the intermediate stage and one from last stage. All other necessary details are provided at the caption of each figure.
 
Selective wetting interaction between A and the particles causes the preferred phase to form a dense enriched/encapsulated layer around the particles. This can be referred as wetting induced primary phase separation which happens at early limes. This causes a composition partitioning in the matrix and a depleted domain of preferred phase results in the particle free region. A rich phase appears in that region at intermediate times due to secondary phase separation. At late times, most of particles are bridged by a interconnected narrow pathways of A rich phase. However, system with higher particle radius (16) does not show this bridge structure. Instead for the case of smaller volume fraction, a strictured interconnected network of A rich phase forms surrounding all the particles in a particular particle dense region.  For higher volume fraction, the late times morphology appears to be as irregular domains of preferred phase dispersed in the matrix and there also a thin wetting layer of minor phase survives about the particle. 
\newpage
\subsection{$A_{60}B_{40}$} 
\begin{figure}[H]
\begin{center}
\includegraphics[trim = 50mm 25mm 50mm 25mm, clip, scale=0.5]{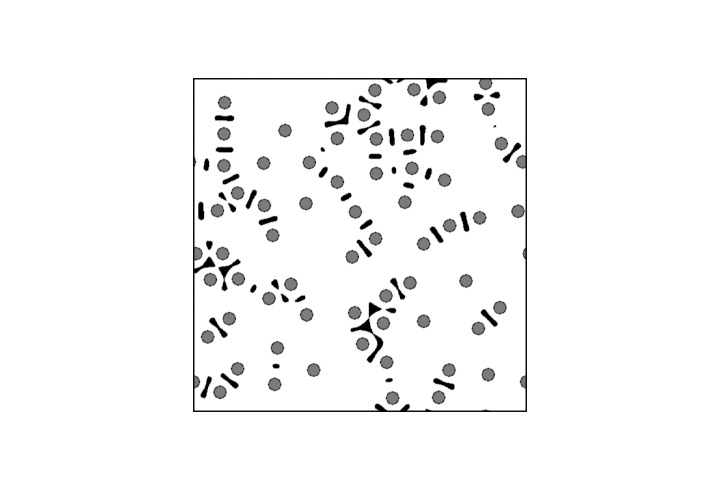}\includegraphics[trim = 50mm 25mm 50mm 25mm, clip,scale=0.5]{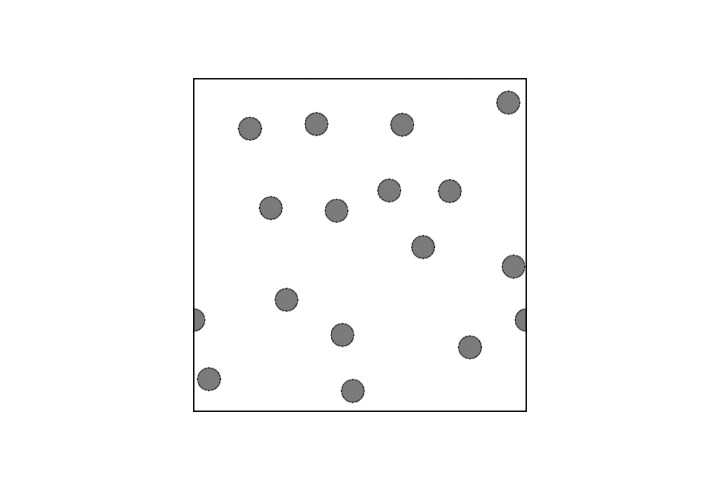}\\
\includegraphics[trim = 50mm 25mm 50mm 25mm, clip, scale=0.5]{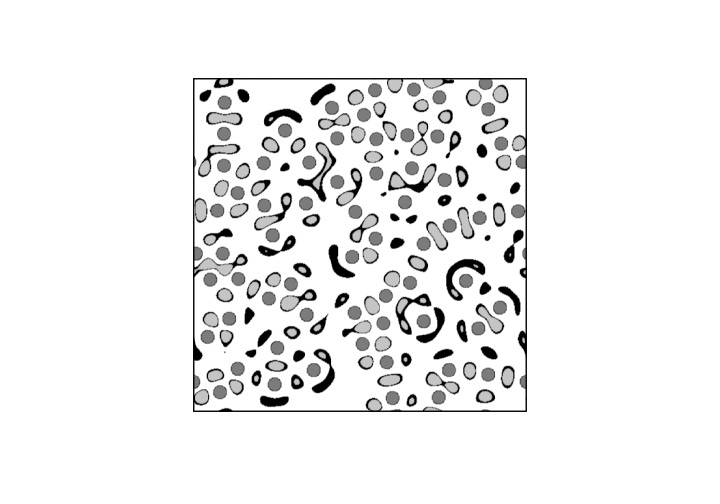}\includegraphics[trim = 50mm 25mm 50mm 25mm, clip,scale=0.5]{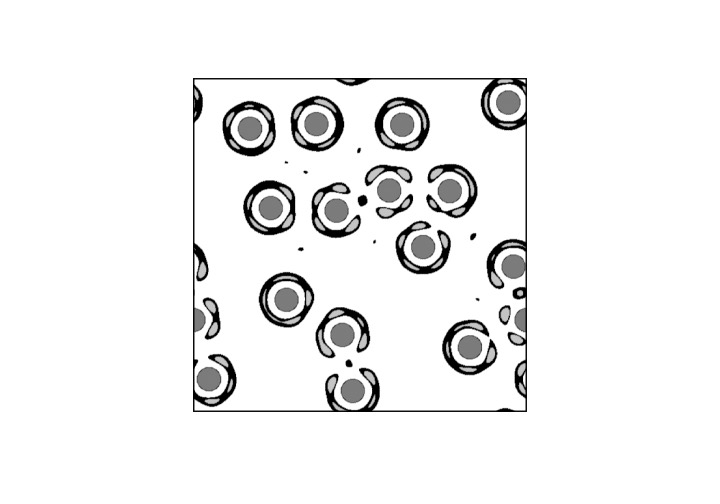}\\
\includegraphics[trim = 50mm 25mm 50mm 25mm, clip, scale=0.5]{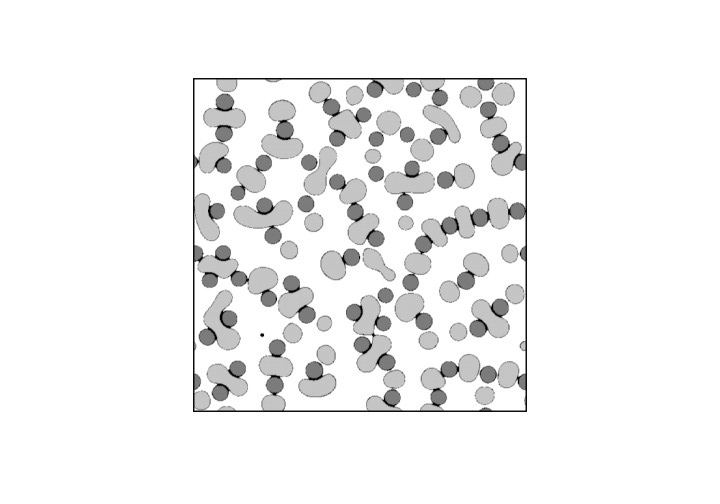}\includegraphics[trim = 50mm 25mm 50mm 25mm, clip,scale=0.5]{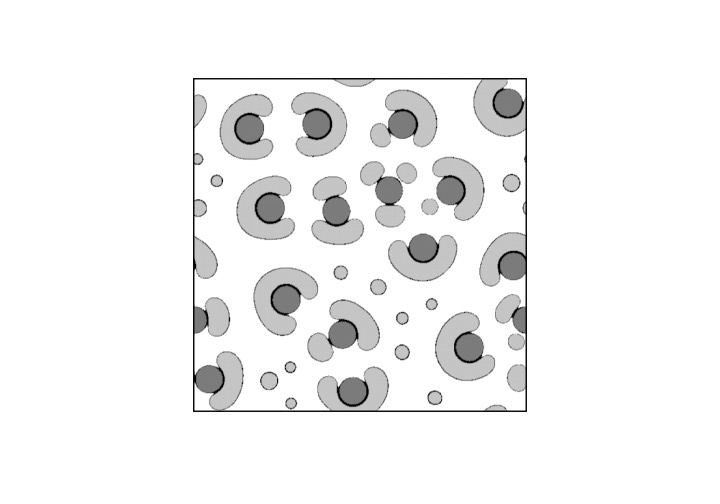}\\
\caption{Microstructures($A_{60}B_{40}$) corresponding to left column is for particle radius of 8 units and right column is for particle radius of 16 units for the same volume fraction (\textbf{5\%}) of particles.The top picture is from some early stage (t = 100 time steps), middle one is of intermediate stage (t = 500 time steps) and bottom one is for late-stage (t = 3000 timesteps). All corresponding microstructures are compared at similar timestep and follow \textbf{system $S_S$}.}\label{o2512_p3}
\end{center}
\end{figure}

\begin{figure}[H]
\begin{center}
\includegraphics[trim = 50mm 25mm 50mm 25mm, clip, scale=0.5]{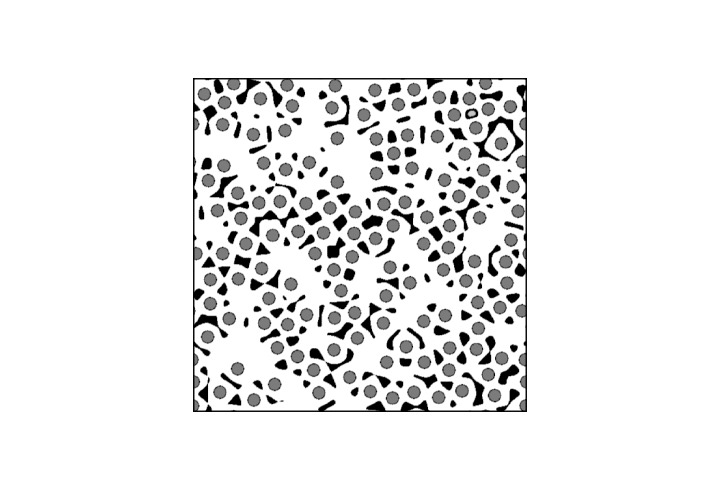}\includegraphics[trim = 50mm 25mm 50mm 25mm, clip,scale=0.5]{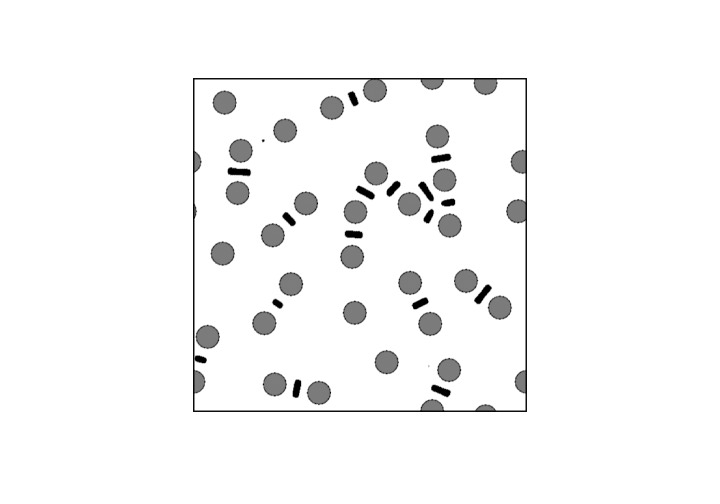}\\
\includegraphics[trim = 50mm 25mm 50mm 25mm, clip, scale=0.5]{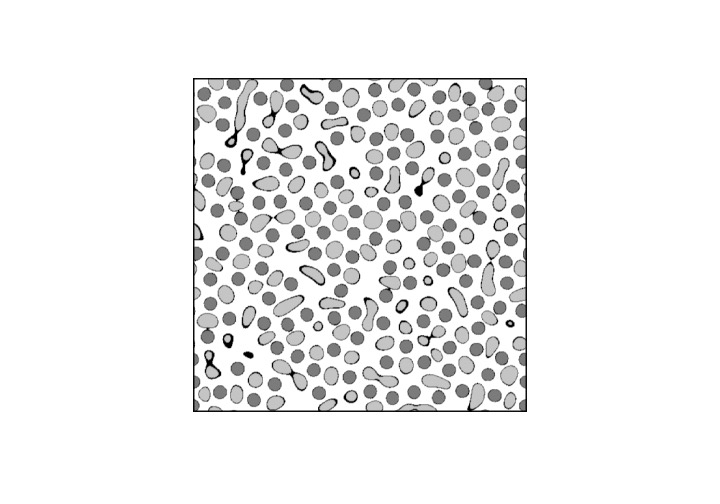}\includegraphics[trim = 50mm 25mm 50mm 25mm, clip,scale=0.5]{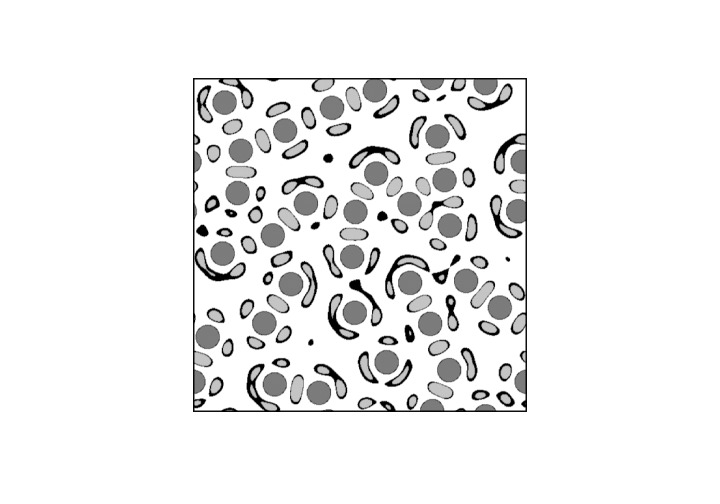}\\
\includegraphics[trim = 50mm 25mm 50mm 25mm, clip, scale=0.5]{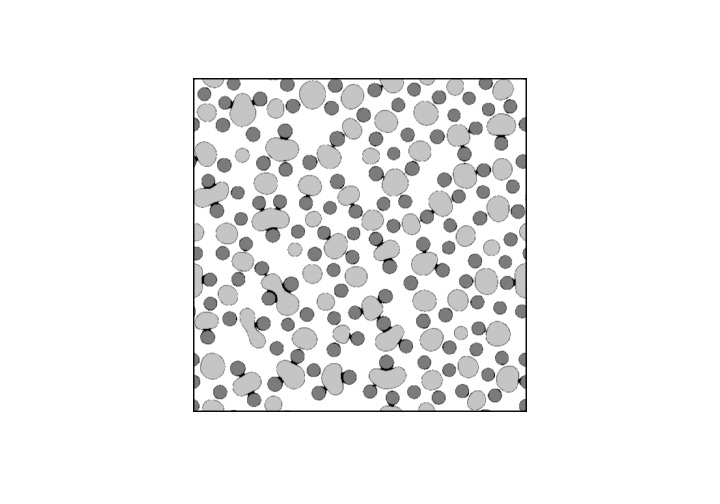}\includegraphics[trim = 50mm 25mm 50mm 25mm, clip,scale=0.5]{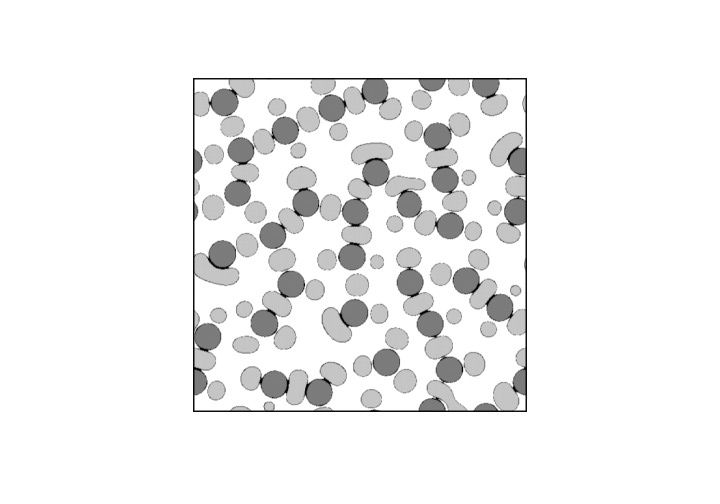}\\
\caption{Microstructures ($A_{60}B_{40}$) corresponding to left column is for particle radius of 8 units and right column is for particle radius of 16 units for the same volume fraction (\textbf{10\%}) of particles.The top picture is of some early stage (t = 100 time steps), middle one is of intermediate stage (t = 500 time steps) and bottom one is of late-stage (t = 3000 timesteps). All corresponding microstructures are compared at similar timestep and follow \textbf{system $S_S$}.}\label{o21012_p3}
\end{center}
\end{figure}

In this case major component (A) wets the particle. For the case volume fraction of particles is 5\%, the corresponding figure is ~\ref{o2512_p3}. Figure ~\ref{o21012_p3} is the case when volume fraction of the particles is 10\%.  We show three snapshots of each case, one from early stage, one from the intermediate stage and one from last stage. All other necessary details are provided at the caption of each figure.

There is no encapsulation layer found about the preferred phase (A). Composition partitioning in the bulk causes a moderately depleted region of A around the wetting layer of particles. In this region B rich (minor) phase nucleates. Sometimes this is referred as "surface induced nucleation". This nucleation further propagates into bulk and B rich phase nucleates in there at still later times. At intermediate to late times the minor phase appears as nearly circular droplets, which are predominantly distributed near to the particle. Smaller radius and higher loading of particles seems to freeze the domains faster at late times.   
\newpage
\subsection{Discussion}
This subsection is an account to address the microstructures obtained when a binary mixture is quenched off-critically ($A_{40}B_{60}$ \& $A_{60}B_{40}$). 

\textbf{More wettable phase is minor phase [Fig. \ref{o1512_p3}, \ref{o11012_p3}]:} system $S_S$ parameters induce a selective interaction of component A and the particles. This eventually leads to a primary phase separation resulting a formation of a uniform A-enriched layer around the particle. Growth of the layer occurs due to the flow of component A from the bulk to the wetting layer. This explains the formation of A-depleted region in the bulk. This is the region where secondary phase separation takes place and a percolating interconnected network of the preferred phase evolves. This morphology is completely different from the case of critically quench microstructures where a ring pattern of alternate layers of A and B forms around the particles or in other words in case of off-critically quench microstructures there is no transient concentric composition waves around the particle~\cite{Lee}. 

At intermediate to late times, the wetting layer seems to feed the growing A-rich domain in the bulk. At late times, the domain size seems to be comparable to the smaller particle (radius = 8) and majority of the particles are located within the interconnected narrow pathways of A-rich phase. Thus, at late times morphology appears to be a bridge pattern~\cite{RTanaka}. The immersed particles in the bridge act as an obstacle to interface motion and stabilize the percolated A- rich domain size. In case of higher density of smaller particles, the bridge accommodates all the particles within it and accordingly domain size is scaled by the immersed particles. In case of bigger particles (radius = 16), the final domain size is not sufficient thick to hold the particles within it. So, A-rich phase become strictured in shape and envelopes a whole region of moderate particle density at every possible location. Thus, at late times more A-B contacts form and system energy is minimized. In case of higher density of larger particles (10\%), particles effectively cut the strictured envelope of A-rich phase, yielding disperse domains of A-rich phase. Here, at late times A-B and A-C contacts are predominant and hence energetically favorable. It is well known that off-critical mixture is supposed to produce a droplet morphology of the minor phase. However, we find a interconnected morphology of the minor phase which is preferred to the particle surface. This phenomena is sometimes referred as inverse percolation to droplet transition (IPCT)~\cite{Ma}.       

\textbf{More wettable phase is major phase [Fig. \ref{o2512_p3}, \ref{o21012_p3}]:} Droplet morphology evolution can be explained as follows. There seems to be double phase separation in this case also. Primary phase separation occurs near to the particle surface at early times and secondary phase separation occurs in the bulk at late stages.

Interphase boundary is a potent site for heterogeneous nucleation~\cite{Porter}. Preferential wetting of component A by the particles causes a lowering of contact angle. The key to reduction of the nucleation barrier is the smaller contact angle. Thus, there is a reduction in local nucleation barrier near the wetted particles. This causes nucleation of minor droplets. This can be referred as "surface induced nucleation"~\cite{Puri_Frisch} and it is the primary one. An interesting explanation of this phenomenon is reported in the paper of Brown and Chakrabarti~\cite{Brown}. The surface induced selective interaction allows the A rich phase to wet around it and expels the B-rich phase to a shorter distance from the wall. This mass transport decreases the nucleation barrier of the B phase because now it is easier to form minor nucleate on a low energy A-B interface.
\\ \\
The optimum shape of the droplets are "two abutted spherical caps" which minimizes the total interfacial energy~\cite{Porter}. As the already formed B droplets grow, composition partitioning occurs in the bulk. Moreover, driving force for nucleation in the bulk, which is far away from wetted particle surface, is very less. That's why kinetics of phase separation in bulk is smaller and the system takes intermediate to late times to form minor droplets in there. At late times the droplets undergo "diffusive coalescence" and attain a nearly spherical shape to reduce the interfacial energy. It is worthy to mention that higher density of smaller particles produce a finest droplet morphology. At late times particles act as an obstacle to interface motion and restricts the growth of domain and droplets. The effect of particle density, particle loading, interfacial effects etc. on the morphology is already described in case of critically quenched microstructres[\ref{cdiss}] and same explanation can be applied to off-critical case also. 

\chapter{Summary and Conclusions}
We have computationally investigated the surface induced phase separation of a binary A:B mixture of critical and off-critical compositions. In our case, surface is provided by spherical particles fixed to the substrate and it exhibits a preferential attraction to one of the components of the binary A:B. To probe this systematically, simulations were carried out following three cases with zero selective interaction, weak selective interaction and strong selective interaction between the particle and preferred component. Off-critical phase separation behavior is studied with strongest selective interaction case only. 
\section*{Symmetric alloy : $A_{50}B_{50}$}
\begin{enumerate}
\item
Phase separation of a critical binary mixture with preferencial attraction results in formation of concentric alternate rings of preferred and non-preferred phases around the particles, whereas absence of preferencial wetting yields no such "target pattern". Such patterns can be referred as core-shell morphology with preferred component as the shell. At late times few core-shell structure still remain, but with different shell (non-preferred component).
\item
The process of engulfing (induced by wetting ) of the preferred component around the particle and subsequent domain coarsening effectively breaks the continuity of domain B. That's why non-wetting domains trapped as isolated islands in a continuous sea of wetting domains and slowing down of domain growth results. This confirms the role of wetting on relaxation of domain growth and generation of an incomplete phase separated morphology.
\item
Particle size and density affects the dynamics of phase separation of a binary mixture. Comparing the microstructures obtained, it is qualitatively true that higher volume fraction and smaller radius of particles are more effective for slowing down the kinetics of phase separation and domain growth.
\item
At late stage, a morphology transition occurs from CW to PW with a surplus of preferred component around the particle, even though the CW morphology is thermodynamically favorable for our set of parameters. The particles act as an obstacle to interfacial motion and thus limits domain growth.
\end{enumerate}
\section*{Off-symmetric alloys : $A_{40}B_{60}$, $A_{60}B_{40}$}
\begin{enumerate}
\item
 When minor component wets the particle, it forms a interconnected network. Whereas when major component wets the surface, no interconnected network of minor component (non-wetting) is observed but droplets of minor components nucleate near the particle surface at the interphase boundary of wetting and non-wetting phases. This is called surface induced nucleation, which predominantly starts near the surface and eventually ends in the bulk at later stage. These droplets coarsen due to interphase diffusion and try to attain a spherical or nearly spherical shape to minimize the interfacial free energy. 
\item
When minor component wets the surface, an enriched layer of preferred component results around the particle. This kind of encapsulation can not be regarded same as "target pattern" because of the fundamental difference in their mechanisms of formation. In case of major component wets the particle, no encapsulation is observed. 
\end{enumerate}
It is very difficult to simulate the experimental conditions exactly by numerical simulation. Variety of variables (physical, chemical, thermal, mechanical) are incorporated during experiments which are very expensive to follow by computation. However, we still believe that our results leads us to a qualitative agreement with the experimental findings. 
\begin{appendices}
\chapter{Thermodynamics of Ternary System}
In case of a heterogeneous ternary system (multicomponent and multiphase) there are three components say A, B, C and A--rich, B--rich and C-- rich phase constitutes the $ \alpha $, $ \beta $ and $ \gamma $ phases respectively. Ternary phase equilibrium is represented as isothermal sections at constant pressure. At equilibrium chemical potential of the components in the existing phases become equal. So, three phase equilibrium in a ternary isothermal phase diagram can be calculated from the following relationships~\cite{lupis} :
\begin{eqnarray}
\mu_A^\alpha = \mu_A^\beta = \mu_A^\gamma\nonumber\\
\mu_B^\alpha = \mu_B^\beta = \mu_B^\gamma\nonumber\\
\mu_C^\alpha = \mu_C^\beta = \mu_C^\gamma
\end{eqnarray}
the above relationships give six individual equations which are as follows : 
\begin{equation}\label{1}\mu_A^\alpha = \mu_A^\beta\end{equation}
\begin{equation}\label{2}\mu_A^\alpha = \mu_A^\gamma\end{equation}
\begin{equation}\label{3}\mu_B^\alpha = \mu_B^\beta\end{equation}
\begin{equation}\label{4}\mu_B^\alpha = \mu_B^\gamma\end{equation}
\begin{equation}\label{5}\mu_C^\alpha = \mu_C^\beta\end{equation}
\begin{equation}\label{6}\mu_C^\alpha = \mu_C^\gamma\end{equation}
Now, in a system comprising three components the chemical potential of each components can be formulated as follows~\cite{lupis} :
\begin{eqnarray}\label{7}
\mu_A &=& f-c_B\frac{\partial f}{\partial c_B}-c_C\frac{\partial f}{\partial c_C}\nonumber\\
\mu_B &=& f+(1-c_B)\frac{\partial f}{\partial c_B}-c_C\frac{\partial f}{\partial c_C}\nonumber\\
\mu_C &=& f-c_B\frac{\partial f}{\partial c_B}+(1-c_C)\frac{\partial f}{\partial c_C}
\end{eqnarray} 
Using the above formalism we can calculate $ \mu_A^\alpha $ and $ \mu_A^\beta $---
\begin{eqnarray}\label{8}
\mu_A^\alpha &=& f^\alpha-c_B^\alpha\frac{\partial f^\alpha}{\partial c_B^\alpha}-c_C^\alpha\frac{\partial f^\alpha}{\partial c_C^\alpha}\nonumber\\
\mu_A^\beta &=& f^\beta - c_B^\beta\frac{\partial f^\beta}{\partial c_B^\beta}-c_C^\beta\frac{\partial f^\beta}{\partial c_C^\beta}
\end{eqnarray}
With the help of ~\ref{8}, equation~\ref{1} can be expanded as :
\begin{equation}\label{f1}
\ln c_A^\alpha-\ln c_A^\beta+\left(c_B^\alpha c_C^\alpha - c_B^\beta c_C^\beta\right)\left(\chi_{AB}-\chi_{BC}+\chi_{AC}\right)+\left(c_B^{\alpha^2}-c_B^{\beta^2}\right)\chi_{AB}+\left(c_C^{\alpha^2}-c_C^{\beta^2}\right)\chi_{AC} = 0
\end{equation} 
Similarly, we can calculate $ \mu_A^\gamma $, $ \mu_B^\alpha $, $ \mu_B^\beta $, $ \mu_B^\gamma $, $ \mu_C^\alpha $, $ \mu_C^\beta $, $ \mu_C^\gamma $ and equations \ref{2}, \ref{3}, \ref{4}, \ref{5}, \ref{6} becomes
 \begin{equation}\label{f2}
\ln c_A^\alpha-\ln c_A^\gamma+\left(c_B^\alpha c_C^\alpha - c_B^\gamma c_C^\gamma\right)\left(\chi_{AB}-\chi_{BC}+\chi_{AC}\right)+\left(c_B^{\alpha^2}-c_B^{\gamma^2}\right)\chi_{AB}+\left(c_C^{\alpha^2}-c_C^{\gamma^2}\right)\chi_{AC} = 0
\end{equation} 
\begin{eqnarray}\label{f3}
\ln c_B^\alpha-\ln c_B^\beta&+&\left(c_B^\alpha c_C^\alpha - c_B^\beta c_C^\beta\right)\left(\chi_{AB}-\chi_{BC}+\chi_{AC}\right)+\chi_{AB}\left[\left(c_A^\alpha-c_A^\beta\right)-\left(c_B^\alpha-c_B^\beta\right)+\left(c_B^{\alpha^2}-c_B^{\beta^2}\right)\right]\nonumber\\
&+&\left(c_C^\alpha-c_C^\beta\right)\left(\chi_{BC}-\chi_{AC}\right)+\left(c_C^{\alpha^2}-c_C^{\beta^2}\right)\chi_{AC} = 0
\end{eqnarray} 
\begin{eqnarray}\label{f4}
\ln c_B^\alpha-\ln c_B^\gamma&+&\left(c_B^\alpha c_C^\alpha - c_B^\gamma c_C^\gamma\right)\left(\chi_{AB}-\chi_{BC}+\chi_{AC}\right)+\chi_{AB}\left[\left(c_A^\alpha-c_A^\gamma\right)-\left(c_B^\alpha-c_B^\gamma\right)+\left(c_B^{\alpha^2}-c_B^{\gamma^2}\right)\right]\nonumber\\
&+&\left(c_C^\alpha-c_C^\gamma\right)\left(\chi_{BC}-\chi_{AC}\right)+\left(c_C^{\alpha^2}-c_C^{\gamma^2}\right)\chi_{AC} = 0
\end{eqnarray} 
\begin{eqnarray}\label{f5}
\ln c_C^\alpha-\ln c_C^\beta&+&\left(c_B^\alpha c_C^\alpha - c_B^\beta c_C^\beta\right)\left(\chi_{AB}-\chi_{BC}+\chi_{AC}\right)+\chi_{AC}\left[\left(c_A^\alpha-c_A^\beta\right)-\left(c_C^\alpha-c_C^\beta\right)+\left(c_C^{\alpha^2}-c_C^{\beta^2}\right)\right]\nonumber\\
&+&\left(c_B^\alpha-c_B^\beta\right)\left(\chi_{BC}-\chi_{AB}\right)+\left(c_B^{\alpha^2}-c_B^{\beta^2}\right)\chi_{AB} = 0
\end{eqnarray}
\begin{eqnarray}\label{f6}
\ln c_C^\alpha-\ln c_C^\gamma&+&\left(c_B^\alpha c_C^\alpha - c_B^\gamma c_C^\gamma\right)\left(\chi_{AB}-\chi_{BC}+\chi_{AC}\right)+\chi_{AB}\left[\left(c_A^\alpha-c_A^\gamma\right)-\left(c_C^\alpha-c_C^\gamma\right)+\left(c_C^{\alpha^2}-c_C^{\gamma^2}\right)\right]\nonumber\\
&+&\left(c_B^\alpha-c_B^\gamma\right)\left(\chi_{BC}-\chi_{AB}\right)+\left(c_B^{\alpha^2}-c_B^{\gamma^2}\right)\chi_{AB} = 0
\end{eqnarray} 

These above equations are solved with respect to following constraints :
\begin{eqnarray}
c_A^\alpha + c_B^\alpha + c_C^\alpha = 1 \nonumber\\
c_A^\beta + c_B^\beta + c_C^\beta = 1 \nonumber\\
c_A^\gamma + c_B^\gamma + c_C^\gamma = 1 
\end{eqnarray}

Thus the number of unknown variables reduced to six. Now it is possible to solve the six non--linear equations with the help of Newton--Raphson Method.  Similarly, two phase equilibrium in a ternary phase diagram can be constructed from the consideration of three component distributed in two phases and chemical potential of the each component in the two phases becomes equal.
\chapter{Interfacial Energy Determination}\label{ied}
Interfacial energy between two co-existing phases, for example $\alpha$--$\beta$, can be calculated following Cahn-Hilliard (CH) formalism~\cite{Cahn}. If a flat one-dimensional system is considered, derivatives of second or higher order are neglected and the cross-sectional area is considered as unity then the equation \ref{che} reduces to the specific interfacial energy ($\sigma$) which is given by the following equation : 
\begin{equation}\label{ge1}
\sigma_{\alpha\beta} = N_v \int_{-\infty}^{\infty}\left[\Delta f(c_i) + \sum_{i=A,B,C}\kappa_{i}\left(\nabla c_i\right)^2 \right]\, dx 
\end{equation}  
where
\begin{equation}\label{ge2}
\Delta f(c) = f(c_i) - \sum_i \mu_i^{\alpha/\beta} \hspace{1cm} i = A, B, C
\end{equation}
$\mu_i^{\alpha\beta}$ is the chemical potential of the species 'i' in phase $\alpha$ or $\beta$ as at equilibrium chemical potential of a given component is the same in both phases. For a system comprising three phases $\alpha $, $\beta $, $\gamma $ -- $ \sigma_{\beta\gamma }$,  $ \sigma_{\alpha\gamma}$ also can be calculated by similar fashion described above.
\end{appendices}

\cleardoublepage
\phantomsection
\bibliographystyle{unsrt}
\bibliography{test}

\end{document}